\documentclass[12pt,preprint]{emulateapj}

\shorttitle{RSGs in NGC 6822 and WLM}
\shortauthors{Levesque \& Massey}

\begin{document}

\title{Spectral Types of Red Supergiants in NGC 6822 and the Wolf-Lundmark-Melotte Galaxy$^{1}$}
\author{Emily M. Levesque$^{2,4}$ and Philip Massey$^{3}$}

\begin{abstract}
We present moderate-resolution spectroscopic observations of red supergiants (RSGs) in the low-metallicity Local Group galaxies NGC 6822 ($Z = 0.4Z_{\odot}$) and Wolf-Lundmark-Melotte (WLM; $Z = 0.1Z_{\odot}$). By combining these observations with reduction techniques for multislit data reduction and flux calibration, we are able to analyze spectroscopic data of 16 RSGs in NGC 6822 and spectrophotometric data of 11 RSGs in WLM. Using these observations we determine spectral types for these massive stars, comparing them to Milky Way and Magellanic Clouds RSGs and thus extending observational evidence of the abundance-dependent shift of RSG spectral types to lower metallicities. In addition, we have uncovered two RSGs with unusually late spectral types (J000158.14-152332.2 in WLM, with a spectral type of M3 I, and J194453.46-144552.6 in NGC 6822, with a spectral type of M4.5 I) and a third RSG (J194449.96-144333.5 in NGC 6822) whose spectral type has varied from a M2.5 in 1997 to a K5 in 2008. All three of these stars could potentially be members of a recently-discovered class of extreme RSG variables.
\end{abstract}

\footnotetext[1]{This paper is based on data gathered with the 6.5 m Magellan telescopes located at Las Campanas, Chile.}
\footnotetext[2]{CASA, Department of Astrophysical and Planetary Sciences, University of Colorado 389-UCB, Boulder, CO 80309, USA; \texttt{Emily.Levesque@colorado.edu}}
\footnotetext[3]{Lowell Observatory, 1400 West Mars Hill Road, Flagstaff, AZ 86001}
\footnotetext[4]{Einstein Fellow}

\section{Introduction}
Red supergiants (RSGs) are the evolved He-burning descendants of moderately massive (10-25M$_{\odot}$) OB stars. Although RSGs are not the most massive or most luminous members of the massive star population, they are the coolest and largest (in physical size), and as such represent an important extreme in the H-R diagram (HRD) that has, in the past, been poorly matched by evolutionary tracks.

The large radii and cool temperatures of RSGs are constrained by the Hayashi limit, the coolest temperature at which a fully convective star would remain hydrostatically stable (Hayashi \& H\-{o}shi 1961). The location of the Hayashi limit is dependent on metallicity, shifting to warmer temperatures (and therefore earlier spectral types) at lower metallicities (Elias et al. 1985, Levesque et al. 2006). However, for many years the placement of the Hayashi limit on the HRD was at odds with the ``observed" locations of RSGs on the HRD, with the stars appearing to be considerably cooler and more luminous than theoretically permitted (Massey 2003, Massey \& Olsen 2003).

In recent years, we have used the new generation of MARCS stellar atmospheres to fit moderate-resolution spectrophotometry of RSGs in the Milky Way (Levesque et al. 2005), the Magellanic Clouds (Levesque et al. 2006), and M31 (Massey et al. 2009). This work has resulted in a significant revision of these stars' effective temperature scales and their position on the H-R diagram. For the Milky Way and the Large Magellanic Cloud, the location of the RSGs is now in excellent agreement with the predictions of the evolutionary tracks. We also see that the distribution of spectral types in these galaxies shifts with metallicity, yielding earlier average RSG spectral types in lower-metallicity galaxies. While this shift is partially due to lower Ti abundances yielding weaker TiO band strengths, it can primarily be attributed to RSG populations in lower-metallicity galaxies having warmer median effective temperatures. This is a consequence of the metallicity-dependent Hayashi limit, in accordance with the predictions of evolutionary models (e.g. Schaerer et al.\ 1993, Charbonnel et al.\ 1993, Maeder \& Meynet 2001, Meynet \& Maeder 2005). The same effect is also observed in the red giant branches of globular clusters: a decrease in the number of heavy elements contributing electrons corresponds to a decrease in opacity and a corresponding increase in the stars' surface temperature (Elias et al.\ 1985, Levesque et al.\ 2006). However, the lowest metallicity RSGs in our sample, from the SMC, remain slightly cooler than the evolutionary models predict, with some RSGs in the sample remaining to the right of the Hayashi track. It is unclear what exactly is causing this discrepancy. What is clear is that massive stellar evolution at low metallicity, particularly in the later stages of these stars' lives, is still poorly understood.

Additionally, three RSGs in the SMC and one in the LMC were found to display unusual variable behavior that has not been previously associated with RSGs, all showing considerable variations in $V$ magnitudes and effective temperatures (and spectral types) (Levesque et al. 2007, Massey et al. 2007a). These stars suffer dramatic physical changes on timescales of months - when they are at their warmest they are also brighter, more luminous, and show an increased amount of extinction. The extinction is characteristic of the effects of circumstellar dust (Massey et al. 2005) and is thought to indicate sporadic dust production from these stars in their coolest states. One parameter these variable stars share is unusually late spectral types; two of these stars, HV 11423 and [M2002] SMC 055188, were previously observed in an M4.5 I state, considerably later and cooler than any other star in the SMC and lying well to the right of the Hayashi limit. The physical explanation for these stars' rapid variations and unusual parameters, and their connection with low-metallicity environments, still remains to be explained.

The metallicities of the RSG populations studied in Levesque et al.\ (2005, 2006) spanned a factor of 5. By comparison, galaxies in the Local Group with known RSG populations span a factor of 15 in metallicity (Massey et al. 2007b). M31 represents the high metallicity extreme, with some studies placing its metallicity as high as 2$\times$ solar (Blair et al.\ 1982, Zaritsky et al.\ 1994; however, see Massey et al.\ 2009 for further discussion of the M31 abundance). At the other extreme lies the Wolf-Lundmark-Melotte (WLM) galaxy, which, at 0.1$\times$ solar (Hodge \& Miller 1995) has the lowest metallicity of any Local Group galaxy currently forming stars (van den Bergh 2000, Massey 2003). Bresolin et al.\ (2006) have previously obtained spectra of 34 early-type supergiants in WLM, but spectra of evolved supergiants in this galaxy have never before been observed (membership of the four suspected G supergiants detected in Bresolin et al.\ 2006 is not known for certain; for further discussion see Massey et al.\ 2007b).

Here we present moderate-resolution spectroscopy of 16 RSGs in NGC 6822 and 11 RSGs in WLM. The stars were selected using {\it UBVRI} from Massey et al. (2007b) and observed using IMACS on the Magellan 6.5m Baade telescope (\S~2). We determined spectral types for the sample (\S~3), comparing these stars to the Milky Way and Magellanic Cloud RSGs and extending observations of the abundance-dependent evolution of RSG spectral types to lower metallicities. We particularly note two RSGs with aberrantly late spectral types, which we consider to be robust variable candidates. Finally, we consider the implications of these new samples and the opportunities they offer for future work (\S~4).

\section{Data Acquisition}
\subsection{Host Galaxies}
WLM was originally discovered by Wolf (1923) and Melotte (1926). It is classified as an Ir IV-V (van den Bergh 1966) and has a star formation rate (SFR) of 0.001 M$_{\odot}$ yr$^{-1}$ (Hodge \& Miller 1995, Hunter \& Elmegreen 2004). Urbaneja et al.\ (2008) determine a distance modulus to WLM of 24.99 $\pm$ 0.10 mag; Gieren et al.\ (2008) find a very similar distance of 24.92 $\pm$ 0.04 based on multiwavelength Cepheid photometry. Massey et al.\ (2007b) obtained broad-band photometry for 7,656 stars seen towards WLM as part of the Local Group Galaxy Survey (LGGS), and determined a total $E(B-V) = 0.07 \pm 0.05$ for this galaxy (foreground $E(B-V) = 0.03$ from Schlegel et al.\ 1998). 

Lee et al.\ (2005) give a nebular abundance for WLM of log(O/H) + 12 = 7.83 $\pm$ 0.06. From an examination of blue supergiants in WLM, Bresolin et al.\ (2006) find an identical abundance of log(O/H) + 12 = 7.83 $\pm$ 0.12. Venn et al.\ (2003) study two blue supergiants in WLM, and find that one of these stars has an extremely high oxygen abundance, with log(O/H) + 12 $\sim$ 8.45; Urbaneja et al.\ (2008) find similar results for this unusual star. Venn et al.\ (2003) speculate that this may be due to spatial chemical inhomogeneities in WLM; however, Lee et al.\ (2005) find no support for this based on their spectroscopic survey of HII regions in WLM. Finally, Urbaneja et al.\ (2008) also found comparable abundances to the Lee et al.\ (2005) nebular metallicity in their study of B and A supergiants in WLM. Venn et al.\ (2003), Lee et al.\ (2005), and Bresolin et al.\ (2006) all determined specific elemental stellar abundances for Fe, Mg, N, and Si. Mg appears to be enhanced in WLM supergiants relative to what is expected from the nebular abundance; Bresolin et al.\ (2006) find enhanced N abundances in their blue supergiant spectra, but speculate that this is a consequence of enhanced rotational mixing effects, an effect which is predicted to be higher in low-metallicity environments (Maeder \& Meynet 2001, Leitherer 2008, Hirschi et al.\ 2008). Finally, Tramper et al.\ (2011) measure an unusually high mass loss rate for a WLM O-star, and speculate that this unexpected result could potentially be attributed to unexpected effects of low-metallicity evolution, such as wind clumping, pulsations, or magnetic fields.

NGC 6822, originally discovered by Barnard (1885), is a barred dwarf irregular galaxy, also classified as an Ir IV-V (van den Bergh et al.\ 2000; Melotte 1926 notes the morphological similarities between NGC 6822 and WLM). This galaxy has an SFR of 0.01 M$_{\odot}$ yr$^{-1}$ (Hunter \& Elmegreen 2004). van den Bergh (2000) gives a distance modulus to this galaxy of 23.45 $\pm$ 0.08 (0.50 Mpc); Pietrzy\`{n}ski et al.\ (2004) find a slightly lower distance modulus of 23.34 $\pm$ 0.09. Massey et al.\ (2007b) obtained broad-band photometry for 51,877 stars seen towards NGC 6822 and calculate a total $E(B-V) = 0.25 \pm 0.02$ for this galaxy, a value which is dominated by the foreground $E(B-V) = 0.22$ (Schlegel et al.\ 1998; it should also be noted than NGC 6822 is at a much lower galactic latitude than WLM, leading to a much higher number of foreground contaminants).

Pagel et al.\ (1980) originally determined a nebular metallicity of log(O/H) + 12 = 8.25 $\pm$ 0.07 for NGC 6822 based on observations of seven different HII regions. More recently, Lee et al.\ (2006) determine direct oxygen abundances from 5 HII regions and calculate a lower log(O/H) + 12 = 8.11 $\pm$ 0.1; however, this abundance is restricted to HII regions with detections of the auroral [OIII] $\lambda$4363 line, suggesting that the survey may be biased towards lower-metallicity HII regions where the feature is strong enough to be detected (e.g. Garnett et al.\ 2004). Lee et al.\ (2006) also find no clear evidence for any abundance gradient in the galaxy. Abundance studies of individual supergiants in this host are limited. Muschielok et al.\ (1999) find an overall [Fe/H] = $-0.5 \pm 0.2$ dex based on spectra of three NGC 6822 B supergiants. Venn et al.\ (2001) determined a metallicity of log(O/H) + 12 = 8.37 $\pm$ 0.21 based on high-resolution spectroscopic studies of two A supergiants in NGC 6822, and find [Fe/H] = $-0.49 \pm 0.22$, in agreement with the Muschielok et al.\ (1999) abundance and giving a metallicity for NGC 6822 that is slightly higher than that of the SMC. However, most recent work in NGC 6822 has focused on the red giant and asymptotic giant branch stellar populations (e.g. Cioni \& Habing 2005, Kang et al.\ 2006, Groenewegen et al.\ 2009); Cioni \& Habing (2005) note that it is difficult to separate RSGs from the lower-mass giants in such work.

\subsection{Sample Selection}
The RSGs in our samples were selected using photometry from LGGS. For both of these hosts, Massey et al.\ (2007b) originally plotted color-magnitude diagrams ($B-V$ vs. $V$) and statistically ``cleaned" the initial data to eliminate foreground dwarfs. Strong RSG populations were visible for both galaxies in this initial analysis.

Massey (1998) originally identified a sample of RSGs in NGC 6822 (as well as samples in M31 and M33) using the two-color method, applying a cut-off line in the $V-R$ vs. $B-V$ diagram that separates RSGs (above the line) from foreground dwarf contaminants (below the line) due to surface gravity effects. To select samples of RSG candidates from the LGGS data for our observations, we began by selecting stars with $V-R > 0.6$ and $V \le 20$, restricting the sample to red stars bright enough to be observable with the Baade 6.5m telescope in exposure times of as little as $\sim$2 hours. With these criteria in place, the initial sample selection was done using the Massey (1998) two-color method, applying a cutoff line of $B-V = 1.25 \times (V-R) + 0.45$ in $V-R$ vs. $B-V$ diagrams for both galaxies (Figure 1). For WLM this yielded an initial list of 18 potential RSG targets. Upon closer examination, one of these targets was found to be a galaxy in {\it BVR} images, and one more was removed due to crowding issues, yielding a final list of 16 RSGs to be observed in WLM. For NGC 6822, the $V$, $V-R$, and $B-V$ criteria produced an initial list of 182 potential RSG targets. After removing 46 targets due to crowding issues, we were left with a final list of 136 RSG to observe in NGC 6822.

Even with foreground dwarfs eliminated, red giants in the halo of the Milky Way can also potentially contaminate samples of RSGs in Local Group galaxies (e.g. Levesque et al.\ 2007, Massey et al.\ 2009). For NGC 6822 and WLM, we use the Besan\c{c}on models of the Milky Way (Robin et al.\ 2003)\footnotemark \footnotetext{Available at http://model.obs-besancon.fr/} to generate the expected number of stars as a function of magnitude and color in the square degree of sky centered on each galaxy. After applying the magnitude and color cuts used to select our RSG sample, we find only one remaining foreground contaminant star in each field, corresponding to a low likelihood of halo giant contamination in each case - 6\% for the WLM and 1\% for NGC 6822. When applied to our final sample size in each galaxy, we therefore conclude that all of the stars selected here are bona fide RSG members of NGC 6822 or WLM. 

\subsection{Observations and Data Reduction}
The RSGs were observed using IMACS (Dressler et al.\ 2011) on the Magellan Baade 6.5-meter telescope at Las Campanas Observatory on 07-08 August 2008. Conditions were clear on the first night, with sporadic high cirrus present on the second night; seeing on both nights was 0.5-0.6". The observations were obtained using five multi-object slitlet masks, three for WLM and two for NGC 6822, with slit widths of 1.2". These masks permitted us to observe all 16 WLM RSGs and 67 of the NGC 6822 RSGs. While all 16 WLM RSGs could have been included in two masks, three were advantageous as multiple orientations allowed us to account for the potential effects of crowding. The IMACS f/4 camera gave us a field of view of 15' $\times$ 15', which is well-matched to the size of the galaxies. Using the 300 line mm$^{-1}$ grating, we were able to obtain spectra with a resolution of 10\AA\ with an exposure time of 3 hours on each mask. The three WLM masks were observed at average airmasses of 1.05, 1.25, and 1.38, while the two NGC 6822 masks were observed at average airmasses of 1.37 and 1.39. Due to the fixed orientation required for multislit masks, our observations were not taken at the parallactic angle; however, IMACS has an Atmospheric Dispersion Corrector (ADC) that compensates for atmospheric differential refraction.

We required continuous spectral coverage between $\sim$5000\AA\ and 8000\AA\ for our analyses. This initially posed a problem due to the configuration of the IMACS Mosaic1 detector at f/4: the detector is an 8192 $\times$ 8192 CCD camera which consists of eight thinned 2K $\times$ 4K $\times$ 15$\mu$ CCD chips, arranged in a 2$\times$4 grid with gaps of $\sim$62 pixels (12.4") between the individual chips. With one spectrum extending across four of these chips, at any single grating tilt sections of the spectrum would fall into these gaps between the CCD chips (see Figure 2 for an example of our raw multislit spectroscopic data from Mosaic1). This was avoided by observing each mask at two different grating tilts, 5.6 degrees (central wavelength of 6015\AA) and 5.8 degrees (central wavelength of 6225\AA), in order to ensure complete spectral coverage for each spectrum.

We reduced the data using IRAF\footnotemark\footnotetext{IRAF is distributed by NOAO, which is operated by AURA, Inc., under cooperative agreement with the NSF.}, rather than the Carnegie Observatories System for MultiObject Spectroscopy (COSMOS) data reduction package for IMACS. This permitted us to take a careful step-by-step approach to the delicate task of flux-calibrating multislit spectroscopic observations across the individual CCD chips of the Mosaic1 detector. We observed several spectrophotometric standards for each mask (LTT1020, LTT7379, LTT9239, and CD-34 324; Hamuy et al.\ 1992, 1994), placing each standard in central slits of our masks on both the upper and lower rows of CCD chips and observing each at both grating tilts. To ensure the best possible flux calibration, we then reduced the data for each individual chip separately, generating a specific sensitivity function (using IRAF's \texttt{sensfunc} task in the \texttt{onedspec} package) derived from the portion of the spectrophotometric standard spectrum that fell on the chip. The RSG spectra on each chip were then calibrated using this sensitivity function and the \texttt{calibrate} function in IRAF, and finally the individual calibrated components of each RSG spectrum were median-combined using IRAF's \texttt{scombine} task to generate one complete stellar spectrum. We note that there are minor ``plateaus" in the stellar spectra at $\sim$6000\AA-6300\AA\ and $\sim$7600\AA-7800\AA, which correspond to the gaps in the CCD mosaic and do not impact our determination of the RSG spectral types.

The main difficulty with flux calibrating the data in this manner arose when considering the horizontal (dispersion axis) geometry of the multislit masks. With multiple slits placed at a variety of positions relative to the grating, the wavelength coverage of each individual spectrum varied slightly as a result. In the case of WLM, where the geometry of the galaxy is narrow in the east-west dimension, this did not pose a significant problem - the wavelength coverage of the RSGs and the spectrophotometric standards were comparable as a whole and allowed us to flux-calibrate the key 5000\AA-8000\AA\ region in all of the spectra. However, for the more extended east-west geometry of NGC 6822 this effect was considerable, and the spectral coverage of the spectrophotometric standards relative to the RSGs was insufficient. In Figure 2 we present the raw data from Mosaic1 for one of our NGC 6822 multislit masks; the bright sky lines illustrate the wide range in wavelength coverage spanned by the 34 spectra observed with this mask. As a result, we were able to flux calibrate the WLM RSGs, but not the NGC 6822 sample, which we leave in normalized relative flux units for the remainder of our analyses. For the WLM RSGs, 6 of the stars had poor S/N or poor wavelength coverage; however, we were able to successfully flux calibrate the remaining 11 spectra. Based on our reduction and analyses, we recommend that any future studies seeking to flux-calibrate multislit data should take particular care to observe spectrophotometric standards at multiple positions along the dispersion axis of a slitmask.

Once the flux calibration for our WLM RSGs was complete, we calculated approximate (monochromatic) $V$ band magnitudes based on the stars' 5556\AA\ fluxes, and compared these to the $V$ magnitudes determined from the LGGS data (Massey et al. 2007b). We found that all of our calibrations for each mask were consistently bright by uniform amounts, with standard deviations of 0.1 mag. This suggested that the disparity we were seeing was simply a grey shift in the spectrum. To confirm this we had to match the fluxes we determined for our flux-calibrated spectrophotometric standards, across a full range of bandwidths ranging from 5000\AA\ to 8000\AA. Comparing our measured fluxes to the standard star files in IRAF's \texttt{onedstds} directory, we found that our flux-calibrated spectra agreed excellently with the measured values, differing by small uniform amounts with standard deviations of several hundredths of a magnitude. Based on this, we can confidently claim that the disagreement between the $V$ fluxes in our spectra and the $V$ fluxes in the LGGS data are the result of an incorrect greyshift, rather than a problem with the flux calibration as a whole. We correct for this by manually greyshifting the observations to agree with the fluxes from LGGS. As a result, our observations cannot be used to determine photometric data for these stars, and the spectral coverage and flux precision is insufficient for stellar atmosphere model fitting. However, the overall shape of the stellar continuum has been properly determined and preserved, and can be used in comparisons with other RSG spectra to determine TiO band depths and spectral types.

\section{Spectral Types}
Our reduced NGC 6822 and WLM RSG stellar spectra were compared to observations of spectral standards from Morgan \& Keenan (1973) by Levesque et al.\ (2005). Since this work originally found that several of the spectral ``standards" had to be reclassified, we also supplement our comparison spectra with additional RSGs from Levesque et al.\ (2005, 2006), to ensure consistency in our spectral types.

For stars with late K or M spectral types, spectral types are based on the strengths of the TiO bands (which themselves are sensitive to physical properties such as temperature and metallicity), with stronger bands corresponding to later spectral types. Specifically, the types are primarily determined from the 6158\AA, 6658\AA, and 7054\AA\ bands following Jaschek \& Jaschek (1990), with TiO bands further in the blue (5167\AA, 5448\AA, 5847\AA ) serving as secondary confirmations of the quality of the fit.

For early-K type stars, this classification is considerably more challenging. Typically, we base these spectral types on the strength of the G band and the Ca I $\lambda$4226 line, since at this wavelength coverage and spectral resolution ($\sim$10\AA) the spectrum of a K-type star is nearly featureless. However, these features are not included in our more limited spectral coverage for these RSGs, and we must therefore base our early-K spectral types entirely on the overall goodness of the spectral continuum fit.

For the NGC 6822 RSGs, we normalized our comparison star spectra for use with the non-flux-calibrated data. This allowed us to determine a best fit for stars with good S/N and evidence of TiO bands in their spectra; however, this also imposes a limitation on determining conclusive spectral types for NGC 6822 RSGs with early- and mid-K spectral types, as we cannot base our fits on the overall SED shape. We must also exclude a number of RSGs with poor or noisy spectra where emerging TiO features cannot be confidently distinguished. As a result, we have limited our NGC 6822 spectral types to stars where we can confidently assign a spectral type of K5 I or later, the earliest type where TiO features begin to emerge. This results in a total sample of 16 NGC 6822 RSGs, with a strong bias towards the late-type RSGs. In the case of the WLM RSGs, we were able to fit all 11 of our flux-calibrated spectra with the spectral standards, and could extend our spectral type determinations to the early-K subtypes; however, it should be stressed that these K-type classifications have a lower confidence that the TiO-based late-K and M spectral types. While we can confidently distinguish between ``early" K type stars with no visible TiO features (K0-1 I or K2-3 I) and late-K types where TiO features begin to emerge (K5 I), the distinction between these early-type classifications is subject to the overall effects of reddening and surface gravity on the spectrum, as well as the goodness of the flux calibration.

The spectral types and colors for these stars are given in Table 1; color indices were determined using {\it BVR} photometry from LGGS and $K_s$ photometry from 2MASS, while previous RSG identifications and spectral types are taken from Massey (1998) or targets from Humphreys (1980) cross-referenced by Massey et al.\ (2007b). It should be noted that, while a relation has previously been seen between ($V-K$)$_0$ and $T_{\rm eff}$ (Levesque et al.\ 2005, 2006), we cannot use this relation with our current data. Converting from $V-K_s$ to ($V-K$)$_0$, where ($V-K$)$_0 -=V - (Ks + 0.04) - (0.88 \times A_V)$ following Schlegel et al.\ (1998) and Carpenter (2001), requires a determination of $A_V$ for each star. At present we can only give conservative lower limits of $A_V = 0.09$ for WLM and $A_V = 0.68$ for NGC 6822 based on foreground extinction (Massey et al.\ 2007b). It is well-known that RSGs can display substantial amounts of excess reddening due to circumstellar dust (e.g. Massey et al.\ 2005). In addition, both Lee et al.\ (2005) and Urbaneja et al.\ (2008) have noted evidence of variations in reddening throughout WLM in particular; Urbaneja et al.\ (2008) find $A_V$ values ranging from 0.09 to 0.90 in their B and A supergiant sample. With the uncertain reddening for each RSG in our sample, we are therefore not currently equipped to estimate $T_{\rm eff}$ for the sample. Similarly, determining bolometric luminosities is also not possible with the current data since, in addition to accounting for unknown reddening effects, RSGs have significant $T_{\rm eff}$-dependent bolometric corrections (a 10\% error in $T_{\rm eff}$ corresponds to a factor of 2 error in bolometric luminosity; Kurucz 1992, Massey \& Olsen 2003). In addition to these RSGs, in Table 2 we list the additional NGC 6822 RSGs from our sample whose spectral types could not be determined due to weak TiO features or noisy spectra.

The spectra of RSGs that we were able to assign spectral types to are shown in Figures 3 (WLM) and 4 (NGC 6822) with spectral coverage from $\sim$5000\AA\ - 8000\AA. From these results, two RSGs immediately stand out in our samples as having unusually late spectral types relative to the overall RSG population in their host galaxies: J000158.14-152332.2 in WLM, with a spectral type of M 3 I, and J194453.46-144552.6 in NGC 6822, with a spectral type of M4.5 I. Both of these stars have spectra dominated by extremely strong TiO features. Interestingly, the spectrum of J000158.14-152332.2 also shows H$\alpha$ in emission, and the SED shows evidence of substantial reddening compared to the other WLM RSGs. Both of these properties, along with its relatively late spectral subtype, are also similar to the unusual late-type dust-enshrouded RSG WOH G64 in the LMC (Ohnaka et al.\ 2008, Levesque et al.\ 2009). From our test of halo giant contamination using the Besan\c{c}on models and our elimination of foreground dwarfs using the Massey (1998) two-color method, we feel confident that both of these are extragalactic RSGs.

The extreme nature of these stars' spectral types, and their significance in studies of unusual RSGs, is illustrated in Figure 5. Here we plot histograms of spectral types for RSGs from the Milky Way (Levesque et al.\ 2005), the Large and Small Magellanic Clouds (Levesque et al.\ 2006, 2007, 2009; Massey et al.\ 2007b), and NGC 6822 and WLM (this work). It should be noted that the selection criteria for RSGs across different host galaxies results in a variation of the parameter space that is sampled; while the Milky Way RSGs were selected based on membership in OB associations, the extragalactic RSG samples were selected based on their $V$ magnitudes and colors. In addition, more observations of RSGs in WLM and NGC 6822 are required to improve the completeness of the plot (particularly for early-K supergiants in NGC 6822). Despite these deficiencies, we can clearly see that these samples follow the trend originally described by Elias et al.\ (1985) and Massey \& Olsen (2003), with the average spectral subtype for RSGs shifting to earlier types at lower metallicities as a result of metallicity effects on TiO band strengths and the effective temperature of the Hayashi limit.

In spite of this overall trend, and the expected accompanying restrictions of the Hayashi limit, unusually late-M supergiants are still present in all of the low-metallicity samples. J000158.14-152332.2 and J194453.46-144552.6 are both marked as clear outliers when compared to the other RSGs in their host galaxies. Also labeled are the variable late-type outliers in the Magellanic Clouds (LMC 170452, SMC 046662, SMC 055188, and SMC 050028; Massey et al.\ 2007, Levesque et al.\ 2007) and the unusual dust-enshrouded RSG WOH G64 in the LMC. From this histogram, it is clear that both J000158.14-152332.2 and J194453.46-144552.6 are excellent candidates for follow-up studies searching for variable or dust-enshrouded RSGs in these low-metallicity host galaxies.

In addition, there is one RSG in our NGC 6822 sample, J194449.96-144333.5, whose spectral type was found to be at odds with previous determinations. Massey (1998) assigned this star a spectral type of M2.5, while we assign it a spectral type of K5. In Figure 6 we show the spectrum used in the Massey (1998) classification, taken on 8 September 1997 at the Kitt Peak National Observatory 4-meter telescope, and compare it to the spectrum from this work. From these observations, it is clear that the star's spectral type appears to be variable, with the 1997 spectrum showing much stronger TiO bands. The 1997 spectrum also shows an unusual strong H$\alpha$ absorption line (equivalent width $\sim$6\AA) that is weaker in the 2008 spectrum (equivalent width $\sim$1\AA). For comparison, Figure 7 shows three different spectra of the previously-identified variable HV 11423, an RSG in the SMC. These spectra show very similar variations to our observations of J194449.96-144333.5; the star appears as a K0 I in 2004, an M4 I in 2005, and an M2.5 I in new observations of the star taken with IMACS on Magellan in August of 2010. We therefore suggest that, in addition to J000158.14-152332.2 and J194453.46-144552.6, J194449.96-144333.5 is a prime candidate for follow-up studies examining variable behavior. In addition, this star illustrates that type-variable RSGs cannot necessarily be identified by aberrantly late spectral types in a single epoch of observations, since it is possible that they were simply captured in a early-type state.

\section{Discussion and Future Work}
We observed extragalactic RSG populations in two low-metallicity Local Group galaxies using IMACS. By combining these multislit observations with careful reductions and flux calibrations, we have acquired spectroscopic data for 16 RSGs in NGC 6822 and spectrophotometric data for 11 RSGs in WLM. We have determined spectral types for each of our RSG spectra, and use these data to extend the metallicity-dependent evolution of average RSG spectral types with abundance to new and lower metallicities. In addition, we have uncovered two RSGs, J194453.46-144552.6 and J000158.14-152332.2, with anomalously late spectral types relative to their host populations, and a third RSG, J194449.96-144333.5, that shows evidence of variable behavior from past spectra.

Understanding the impact of metallicity on the observed properties of RSGs is vital to studies of extragalactic massive stars.  The physical properties of RSGs, along with the phenomena driving their mass loss processes, are critical in local studies of supernova progenitors (e.g. Smartt et al.\ 2009, Georgy 2012) and in high-redshift studies of young stellar populations where RSGs are the dominant dust producers (Massey et al.\ 2005). In addition, the trend toward earlier spectral types  poses a potential problem for fully sampling the evolved supergiant populations in lower-metallicity hosts such as NGC 6822, WLM, and the SMC. Any such survey must make an effort to include {\it yellow} supergiants, due to the metallicity-dependent evolutionary effects on the Hayashi limit and the resulting skew towards an evolved supergiant population with slightly bluer colors at lower metallicities as a whole. However, removing foreground contaminants in yellow supergiant samples is challenging and requires detailed kinematic treatments (e.g. Drout et al.\ 2009, Neugent et al.\ 2010).

Finally, the potential addition of two new stars to the sample of variable RSGs is extremely valuable. This is a small and poorly-understood sample of evolved massive stars, one that would benefit from additional observations or regular monitoring programs. New candidates could be detected through photometric observations, due to variations in these stars' color indices, but unambiguous detections and confirmation of variability in stars such as J194449.96-144333.5, J194453.46-144552.6, and J000158.14-152332.2 will require multiple spectroscopic observations. Confirming variable behavior in new candidates, and beginning to understand the physical phenomena driving these stars' behavior, will require additional observations spanning a baseline of several years.

We thank the anonymous referee for their valuable feedback on this manuscript. We also gratefully acknowledge valuable discussions with Nidia Morrell.  We are thankful to the staff of Las Campanas Observatory for their assistance and hospitality during our observations. Magellan telescope time was granted by NOAO, through the Telescope System Instrumentation Program (TSIP). TSIP is funded by NSF. This paper made use of data from the Two Micron All Sky Survey (2MASS), which is a joint project of the University of Massachusetts and the Infrared Processes and Analysis Center, California Institute of Technology, funded by the National Aeronautics and Space Administration and the National Science Foundation. E. M. L. is supported by NASA through Einstein Postdoctoral Fellowship grant number PF0-110075 awarded by the Chandra X-ray Center, which is operated by the Smithsonian Astrophysical Observatory for NASA under contract NAS8-03060. This work was supported by the National Science Foundation through AST-1008020 to P. M.

\begin{deluxetable}{l c c c c c c c}
\tabletypesize{\scriptsize}
\tablewidth{0pc}
\tablenum{1}
\tablecolumns{8}
\tablecaption{\label{tab:EW} WLM and NGC 6822 Red Supergiants}
\tablehead{
\colhead{Star (LGGS)}
& \colhead{Type}
& \colhead{$V$\tablenotemark{a}}
& \colhead{$B-V$\tablenotemark{a}}
& \colhead{$V-R$\tablenotemark{a}} 
& \colhead{$V-K_s$\tablenotemark{a,b}}
& \multicolumn{2}{c}{Previous Work} \\ \cline{7-8}
& \multicolumn{5}{c}{}
& \colhead{Type}
& \colhead{Ref \tablenotemark{c}}
}
\startdata
WLM & & & & & & & \\
\hline 
J000153.17-152813.4 &K0-1 & 19.34 & 1.66 & 0.83 &4.07 &\nodata & \nodata   \\ 
J000156.77-152839.6 &K2-3 & 17.61 & 1.94 & 0.99 & 3.81  & \nodata & \nodata   \\ 
J000156.87-153122.3 &K0-1 & 18.76 & 1.64 & 0.85 & 3.19  & \nodata & \nodata   \\ 
J000157.01-152954.0 &K0-1 & 18.70 & 1.64 & 0.85 & 3.43  & \nodata & \nodata  \\ 
J000157.55-152915.8 &K0-1 & 19.27 & 1.64 & 0.86 & 3.55  & \nodata & \nodata  \\ 
J000157.96-152803.1 &K0-1 & 19.80 & 1.53 & 0.77 &\nodata  & \nodata & \nodata  \\ 
J000158.14-152332.2 &M3   & 19.61 & 1.72 & 0.99 & \nodata  & \nodata & \nodata  \\ 
J000158.74-152245.5 &K0-1 & 18.46 & 1.68 & 0.90 & 3.12  & \nodata & \nodata  \\ 
J000159.61-153059.9 &K2-3 & 18.98 & 1.78 & 0.91 & 3.69  & \nodata & \nodata  \\ 
J000200.81-153115.7 &K0-1 &18.69  & 1.78 & 0.91 & 3.95  & \nodata & \nodata  \\ 
J000203.04-153033.7 &K5   & 18.68 & 1.98 & 1.04 & 4.41  & \nodata & \nodata  \\ 
\hline
NGC 6822 & & & & & & & \\ 
\hline
J194445.76-145221.2 &M1   &17.71 &2.22 &1.21 &4.87   & \nodata & \nodata    \\ %NGCB, RSG9
J194447.56-145215.4 &K5   &19.13 &2.05 &1.12 &4.80     & \nodata & \nodata \\ %NGCB, RSG10 
J194447.81-145052.5 &M1   &18.51 &2.23 &1.29 &5.26    & \nodata & \nodata  \\ %NGCB, RSG14
J194449.96-144333.5 &K5   &18.07 &2.21 &1.29 &5.24    &M2.5 I &1   \\ %NGCA, RSG35 
J194450.12-144637.9 &K5   &19.84 &1.71 &0.90 &\nodata  & \nodata &\nodata  \\ %NGCA, RSG26 
J194450.44-144410.0 &M2   &18.46 &2.00 &1.05 &4.29    & \nodata & \nodata   \\ %NGCA, RSG32 
J194453.46-144540.1 &K5   &19.14 &1.74 &0.97 &4.27    & \nodata & \nodata  \\ %NGCB, RSG28 
J194453.46-144552.6 &M4.5 &18.43 &1.93 &1.05 &4.35    & \nodata & \nodata   \\ %NGCA, RSG28 
J194453.96-144424.3 &K5   &18.19 &2.01 &1.04 &4.46   & \nodata & 1   \\ %NGCA, RSG31
J194454.46-144806.2 &M1   &18.56 &2.00 &1.21 &5.35   & \nodata & \nodata    \\ %NGCB, RSG22 
J194454.54-145127.1 &M0   &17.05 &2.25 &1.19 &4.71  &M0-1 I & 2    \\ %NGCA, RSG15
J194455.70-145155.4 &M0   &16.91 &2.20 &1.17 &4.55  &M1-2 I & 2     \\ %NGCA, RSG13
J194455.93-144719.6 &K5   &19.56 &1.87 &1.04 &4.46    & \nodata & \nodata   \\ %NGCB, RSG23
J194457.31-144920.2 &M1   &17.41 &2.28 &1.21 &4.93    &M1 I &2   \\ %NGCB, RSG20
J194459.86-144515.4 &M1   &16.93 &2.00 &1.00 &4.27    &M0 I &1,2   \\ %NGCA, RSG29
J194503.58-144337.6 &M0   &19.23 &1.91 &1.01 &4.27  & \nodata &  \nodata   \\ %NGCB, RSG34
\hline
\enddata
\tablenotetext{a}{From Massey et al.\ (2007b).}
\tablenotetext{b}{$K_s$ magnitudes are taken from 2MASS where available.}
\tablenotetext{c}{References: 1) Massey (1998). 2) Spectral type from Humphreys (1980), cross-identifications from Massey et al.\ (2007b).}
\end{deluxetable}

\begin{deluxetable}{l c c c c}
\tabletypesize{\scriptsize}
\tablewidth{0pc}
\tablenum{2}
\tablecolumns{5}
\tablecaption{\label{tab:EW} Additional NGC 6822 Red Supergiants}
\tablehead{
\colhead{Star (LGGS)}
& \colhead{$V$\tablenotemark{a}}
& \colhead{$B-V$\tablenotemark{a}}
& \colhead{$V-R$\tablenotemark{a}} 
& \colhead{$V-K_s$\tablenotemark{a,b}}
}
\startdata
J194449.62-145327.3 &19.470  &1.769  &0.957  &4.184 \\ 
J194451.78-144616.7 &19.848  &1.720  &0.895  &\nodata \\ 
J194453.18-145406.5 &19.604  &1.839  &0.982  &4.375 \\ 
J194453.66-144617.9 &19.050  &2.030  &1.169  &4.872 \\    
J194453.68-145245.8 &19.252  &1.816  &0.979  &4.270 \\        
J194454.81-144347.8 &18.050  &2.239  &1.284  &5.324 \\ 
J194454.86-145216.3 &18.768  &1.854  &1.062  &\nodata \\             
J194455.02-144607.1 &19.465  &1.900  &0.993  &4.200 \\            
J194455.19-144344.8 &19.814  &1.652  &0.931  &4.845 \\           
J194455.37-145229.2 &19.272  &1.798  &0.976  &3.941 \\         
J194455.97-145109.4 &19.401  &1.840  &1.013  &4.302 \\  
J194456.77-144927.0 &19.879  &1.655  &0.883  &\nodata \\            
J194457.10-145141.1 &19.950  &1.700  &0.961  &\nodata \\              
J194457.38-145408.0 &18.508  &2.095  &1.114  &4.555 \\             
J194457.69-144953.1 &19.137  &1.857  &1.005  &4.594 \\           
J194457.96-144847.9 &19.991  &1.760  &0.946  &\nodata \\            
J194458.16-144456.0 &18.329  &2.067  &1.126  &4.751 \\             
J194458.17-144527.8 &18.143  &1.924  &0.977  &\nodata \\              
J194458.45-145458.8 &19.553  &1.580  &0.836  &\nodata \\            
J194458.64-145007.4 &19.669  &1.733  &0.912  &\nodata \\               
J194459.22-144955.0 &19.764  &1.730  &0.940  &\nodata \\            
J194459.41-145106.8 &19.877  &1.721  &0.948  &\nodata \\            
J194500.10-144901.7 &19.789  &1.664  &0.930  &4.346 \\             
J194500.32-144104.8 &18.246  &1.999  &1.062  &4.553 \\           
J194500.40-145333.3 &19.674  &1.896  &0.998  &4.158 \\           
J194500.53-144826.5 &18.750  &2.092  &1.238  &5.216 \\            
J194500.91-145234.3 &19.450  &1.788  &0.976  &4.090 \\          
J194501.12-145437.7 &18.979  &1.813  &0.971  &4.239 \\           
J194503.27-144224.5 &19.901  &1.718  &0.927  &\nodata \\             
J194503.87-144247.6 &19.957  &2.060  &1.120  &4.801 \\           
J194504.29-145406.2 &19.456  &1.915  &1.059  &4.571 \\           
J194504.74-145422.8 &19.862  &1.706  &0.926  &4.574 \\           
J194505.74-145547.6 &18.862  &1.892  &1.021  &4.361 \\           
J194505.84-145517.3 &18.935  &1.774  &0.967  &4.281 \\          
J194506.34-145537.0 &19.382  &1.951  &1.041  &4.283 \\          
J194506.35-145135.5 &19.155  &2.076  &1.140  &4.672 \\          
J194507.48-145107.1 &19.677  &1.881  &1.002  &\nodata \\          
J194507.87-145435.9 &19.822  &1.813  &0.943  &\nodata \\              
J194509.55-145344.6 &18.838  &1.880  &0.985  &4.189 \\        
J194515.10-144803.9 &19.263  &1.761  &0.781  &3.919 \\ 
\enddata
\tablenotetext{a}{From Massey et al.\ (2007b).}
\tablenotetext{b}{$K_s$ magnitudes are taken from 2MASS where available.}
\end{deluxetable}

\begin{figure}
\center
\epsscale{0.55}
\plotone{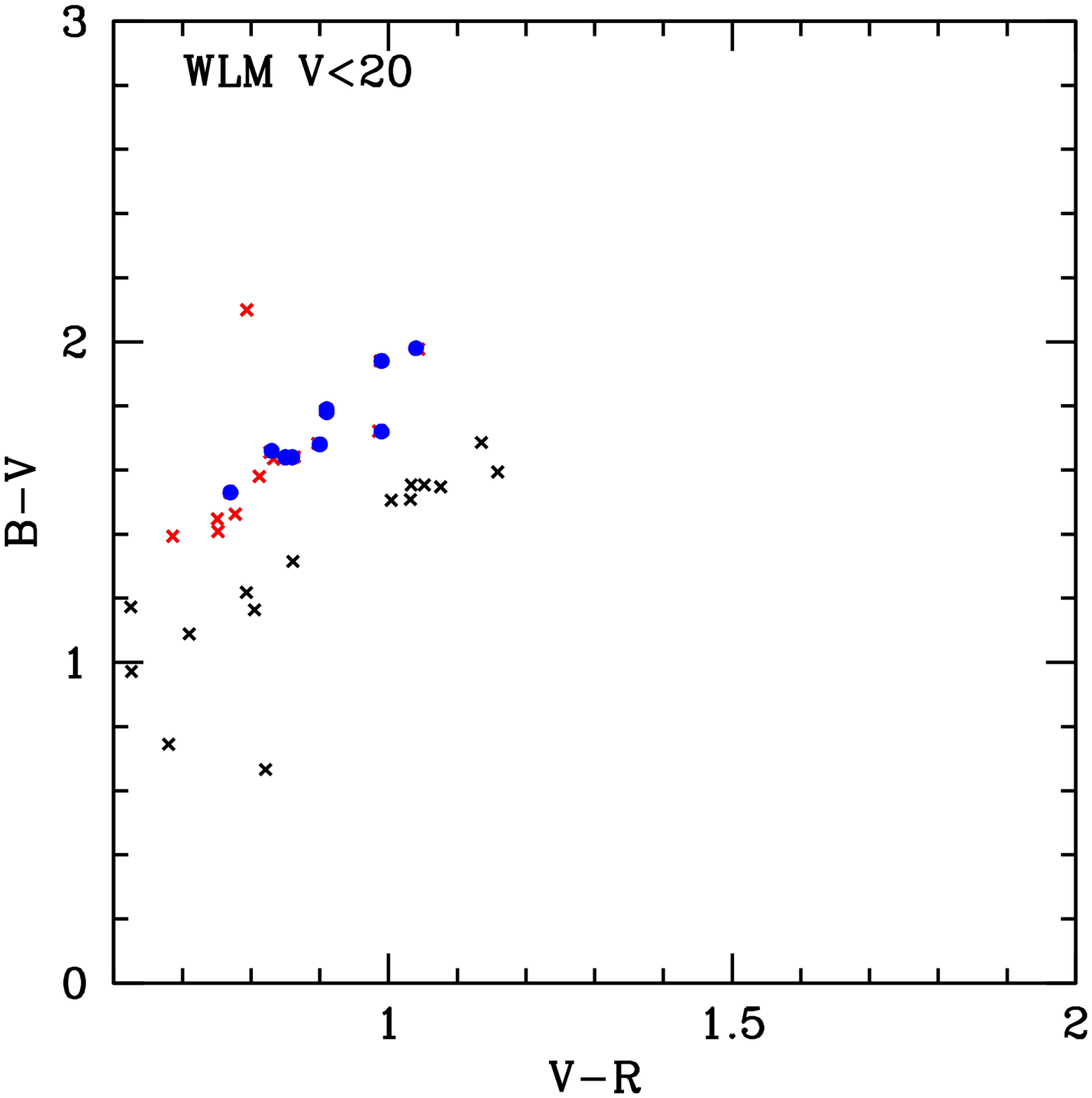}
\plotone{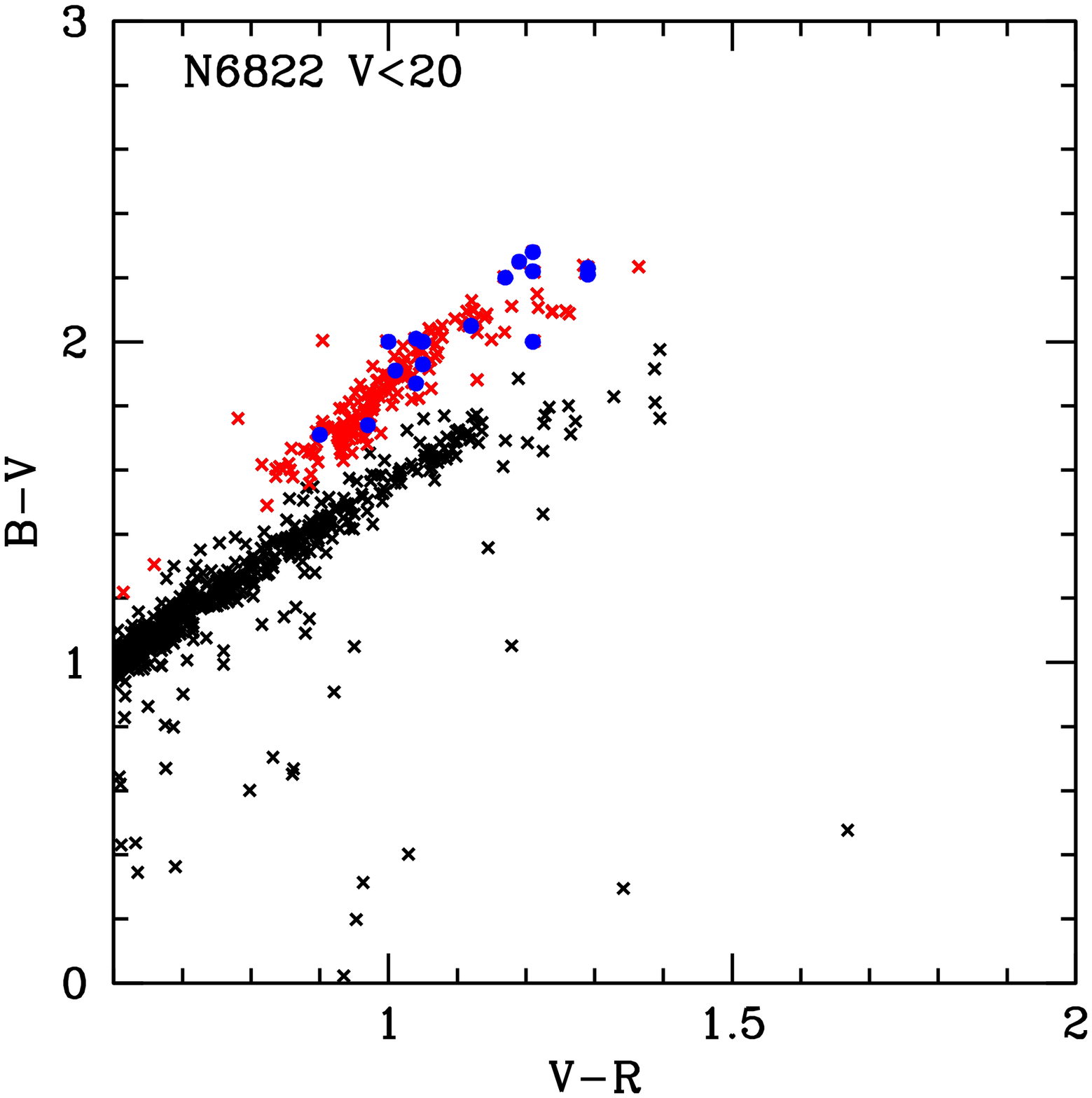}
\caption{$V-R$ vs. $B-V$ plots used to eliminate foreground dwarfs from LGGS photometry centered on WLM (left) and NGC 6822 (right). Presumed RSGs with $V < 20$ are shown as red crosses; foreground dwarfs are plotted as black crosses. Targets with observations presented in this paper are overplotted as blue circles.}
\end{figure}

\begin{figure}
\center
\includegraphics[angle=90,width=16cm]{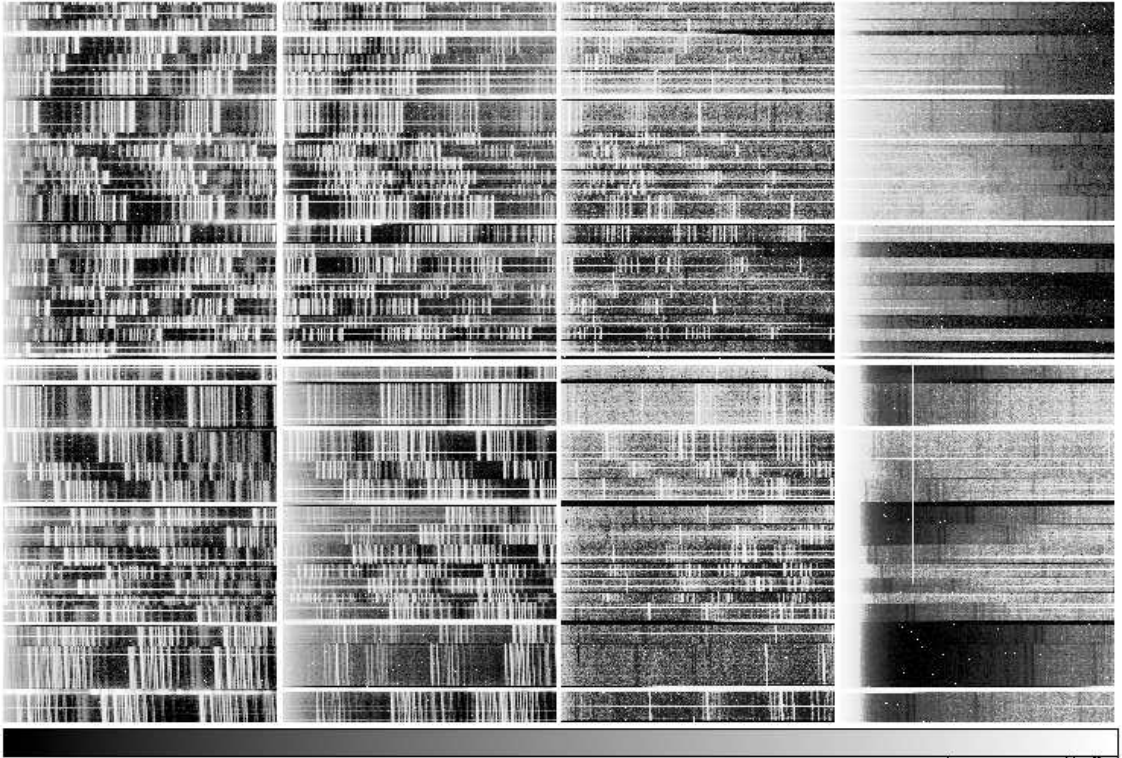}
\vspace{-40pt}
\caption{Reconstructed example of the IMACS Mosaic1 CCD chip geometry, using raw data from our observations of one of our two multislit masks for NGC 6822. Wavelength decreases from left to right in this image. The gaps between the individual CCD chips are shown, illustrating the need for observations at two different grating tilts to ensure complete spectral coverage. The varying x-axis positions of the bright sky lines illustrate the wide range in wavelength coverage for the 34 RSG spectra included on these masks. The chips in the bluest quarter of the CCD (far right) were found to have extremely poor S/N and were not used in our subsequent reductions and analyses.}
\end{figure}

\begin{figure}
\epsscale{0.36}
\plotone{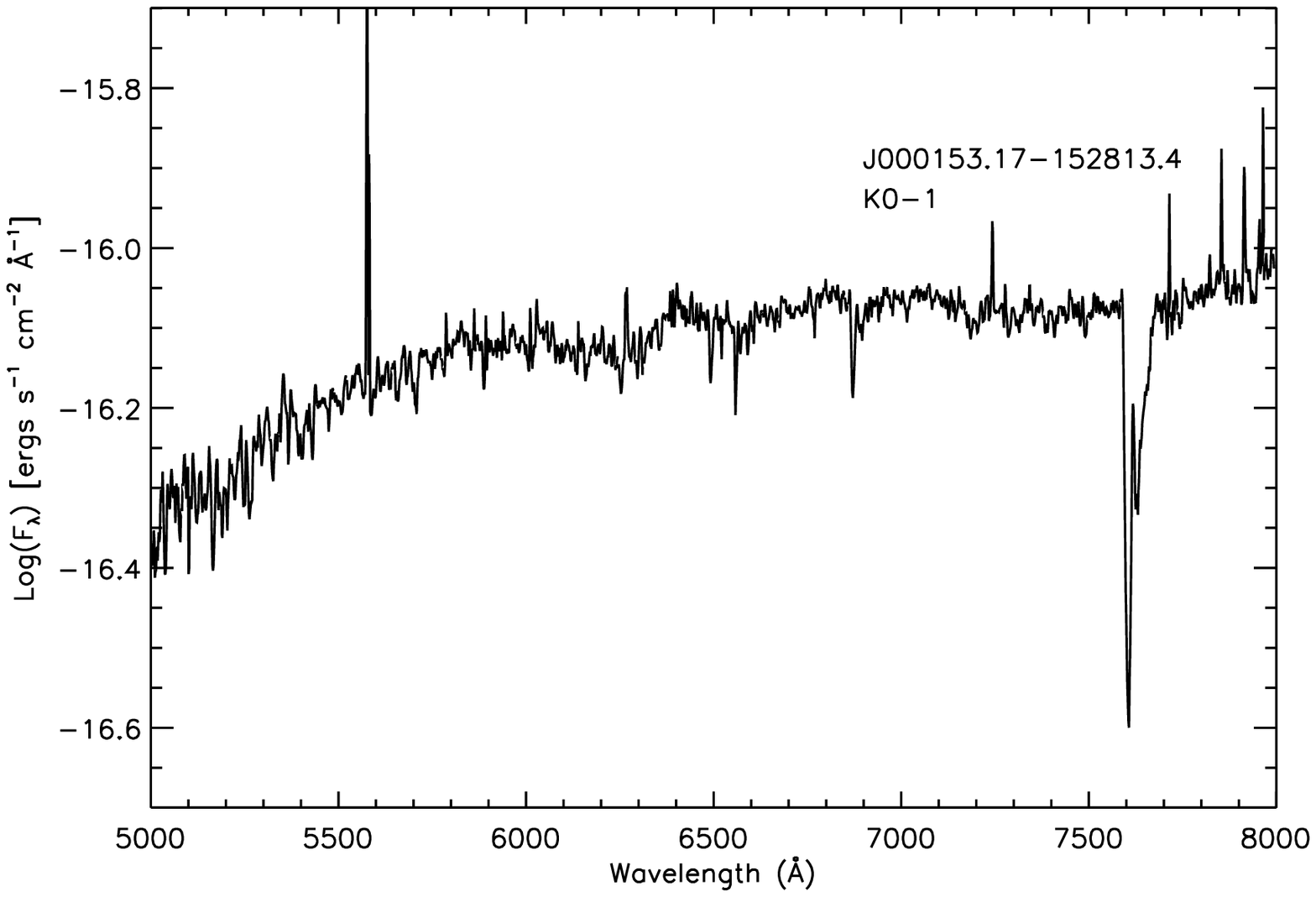}
\plotone{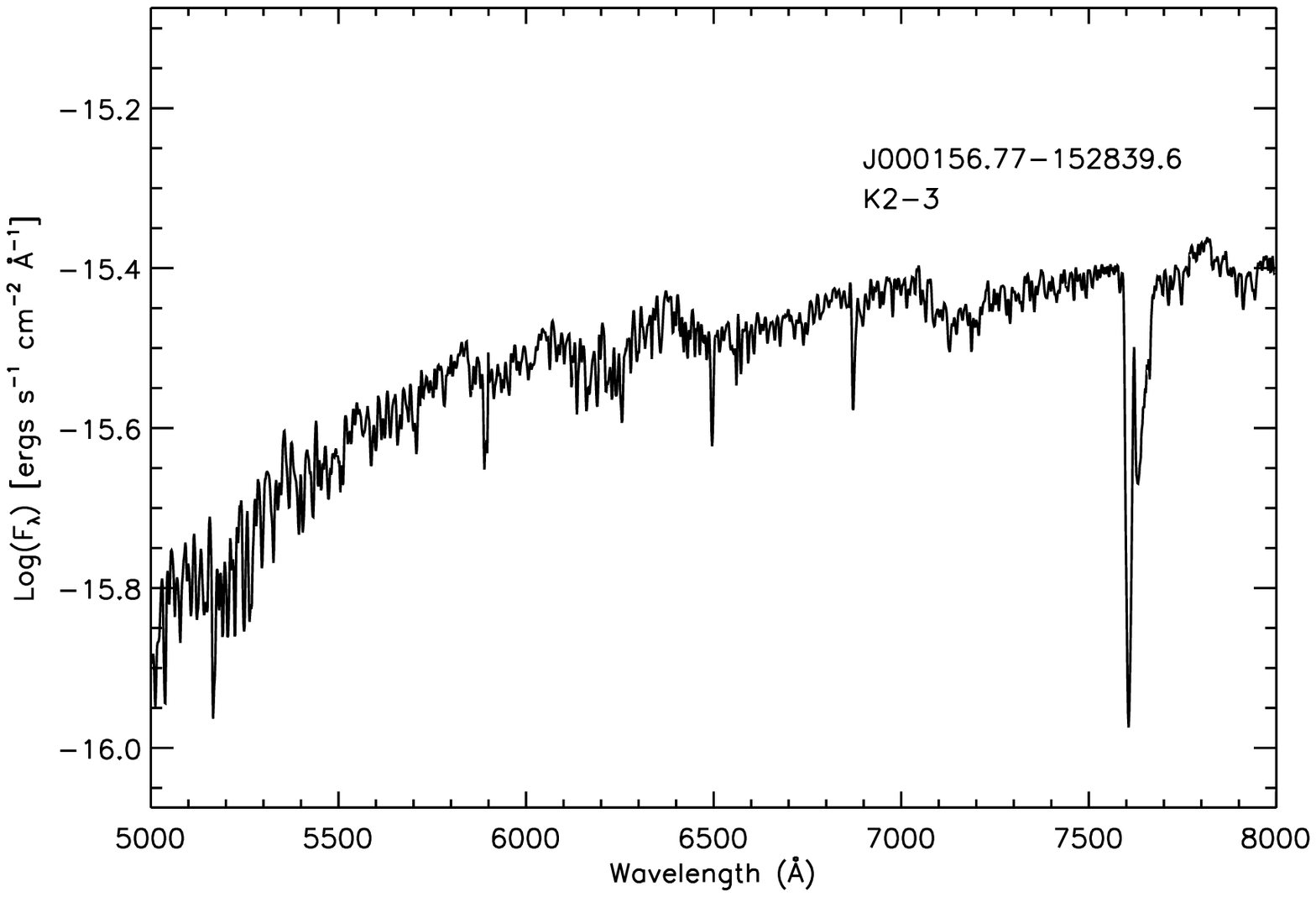}
\plotone{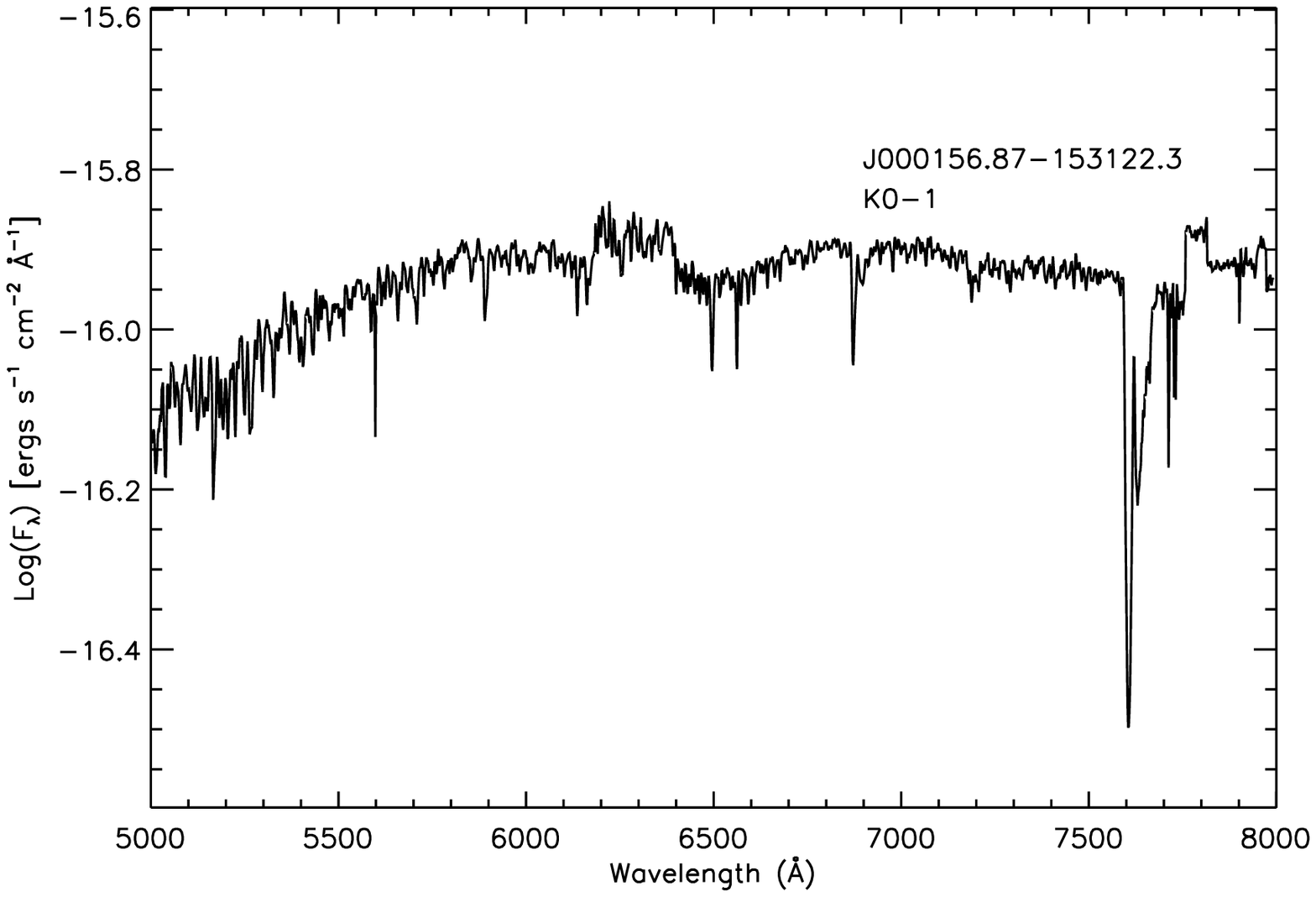}
\plotone{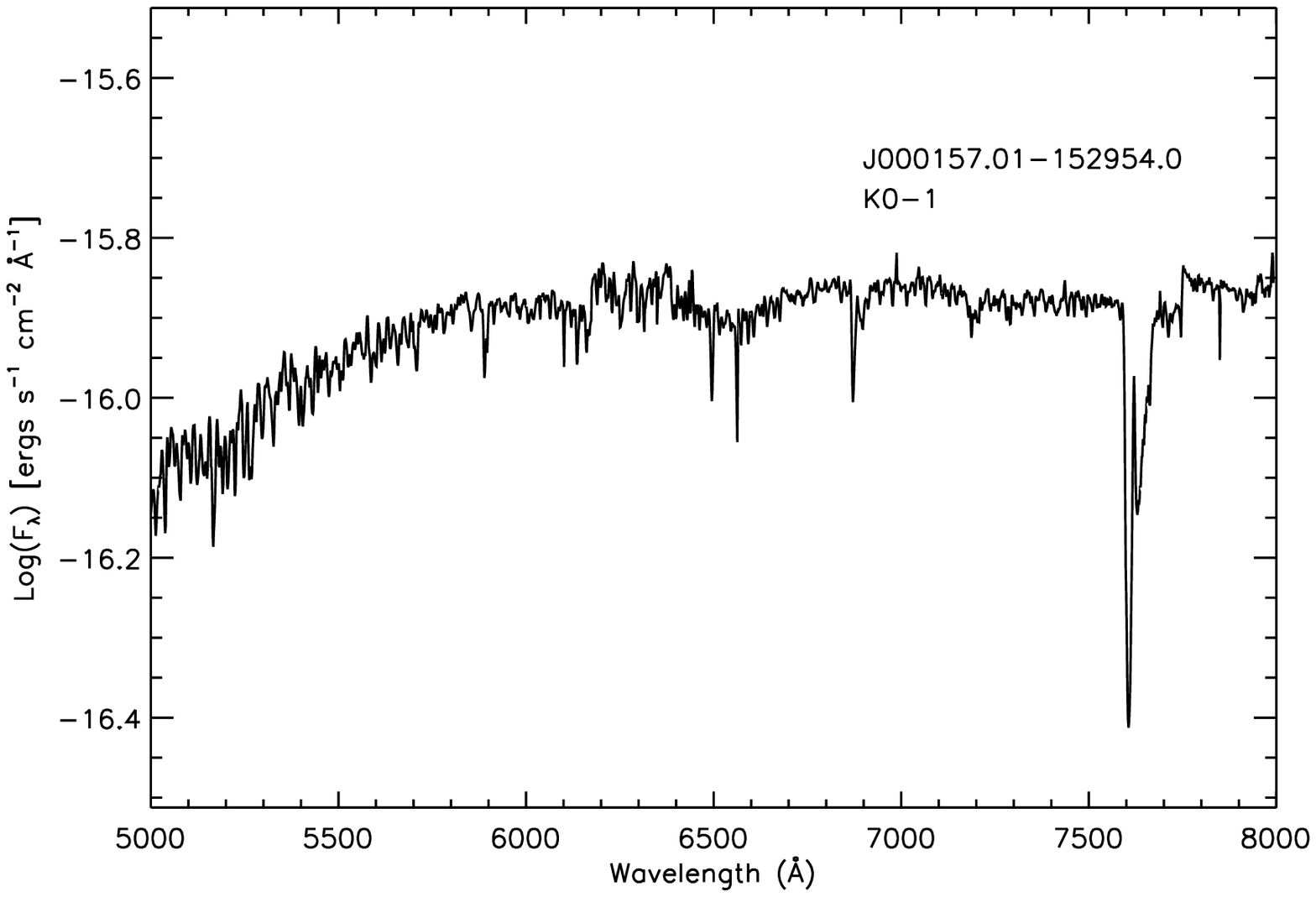}
\plotone{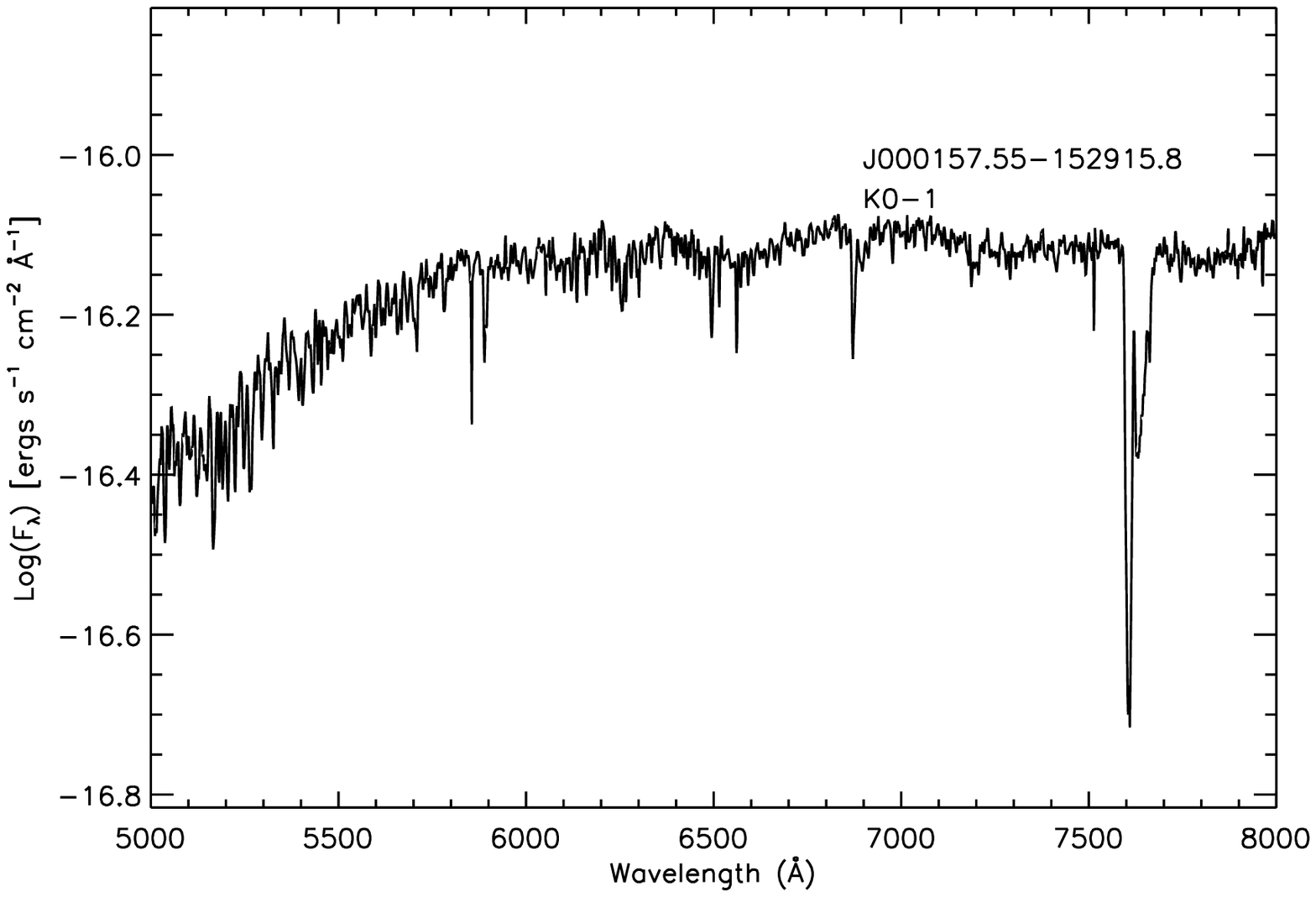}
\plotone{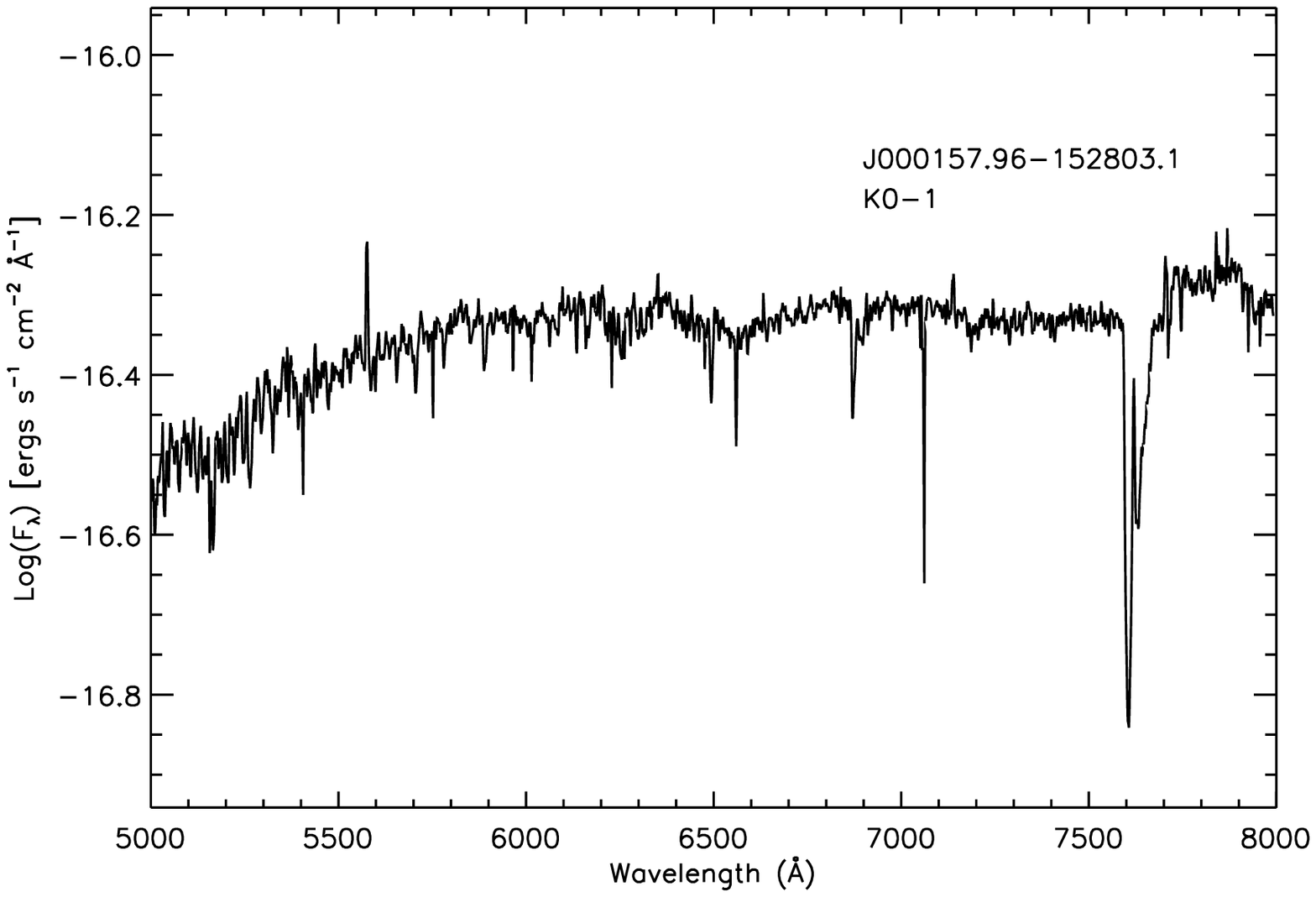}
\plotone{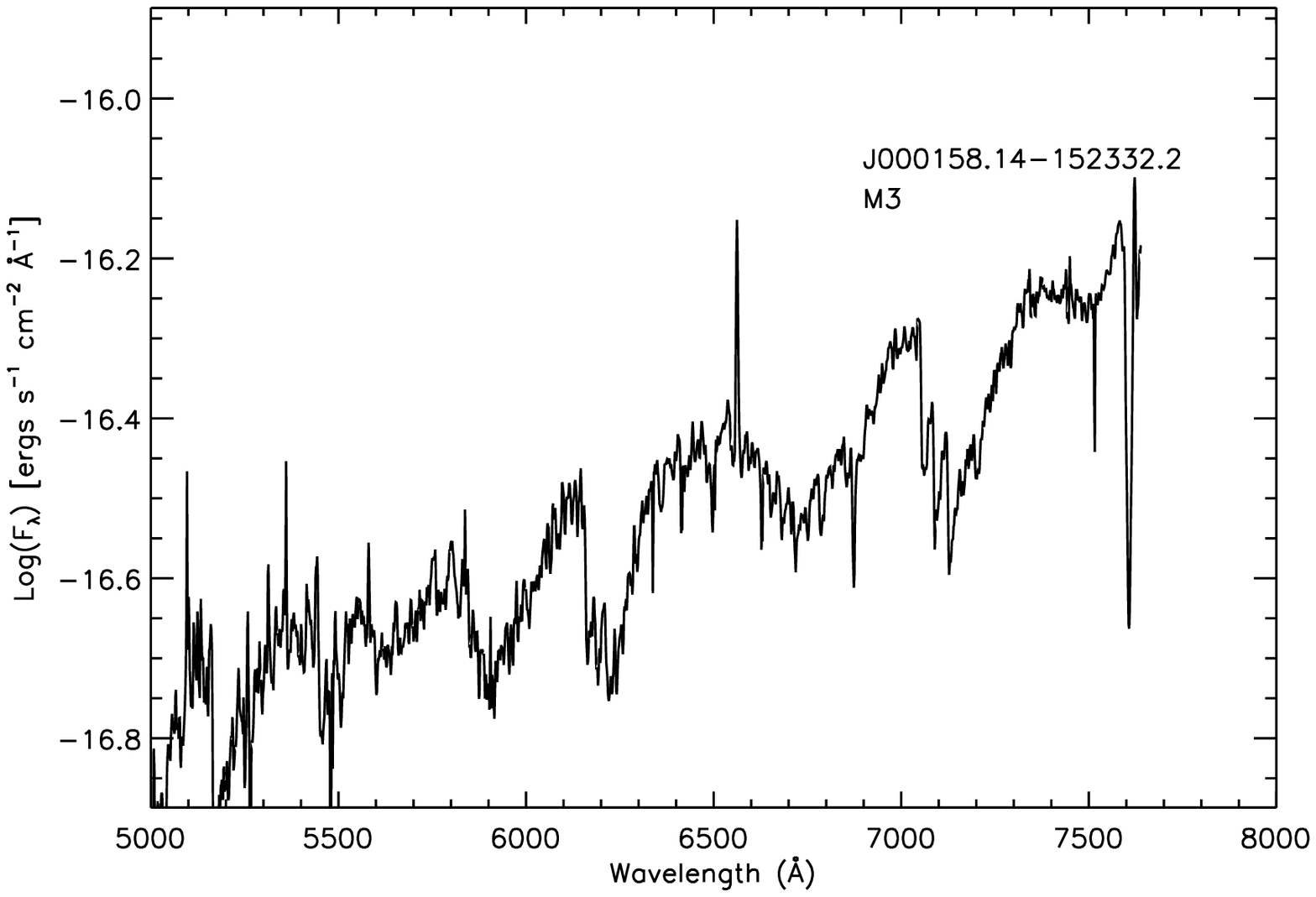}
\plotone{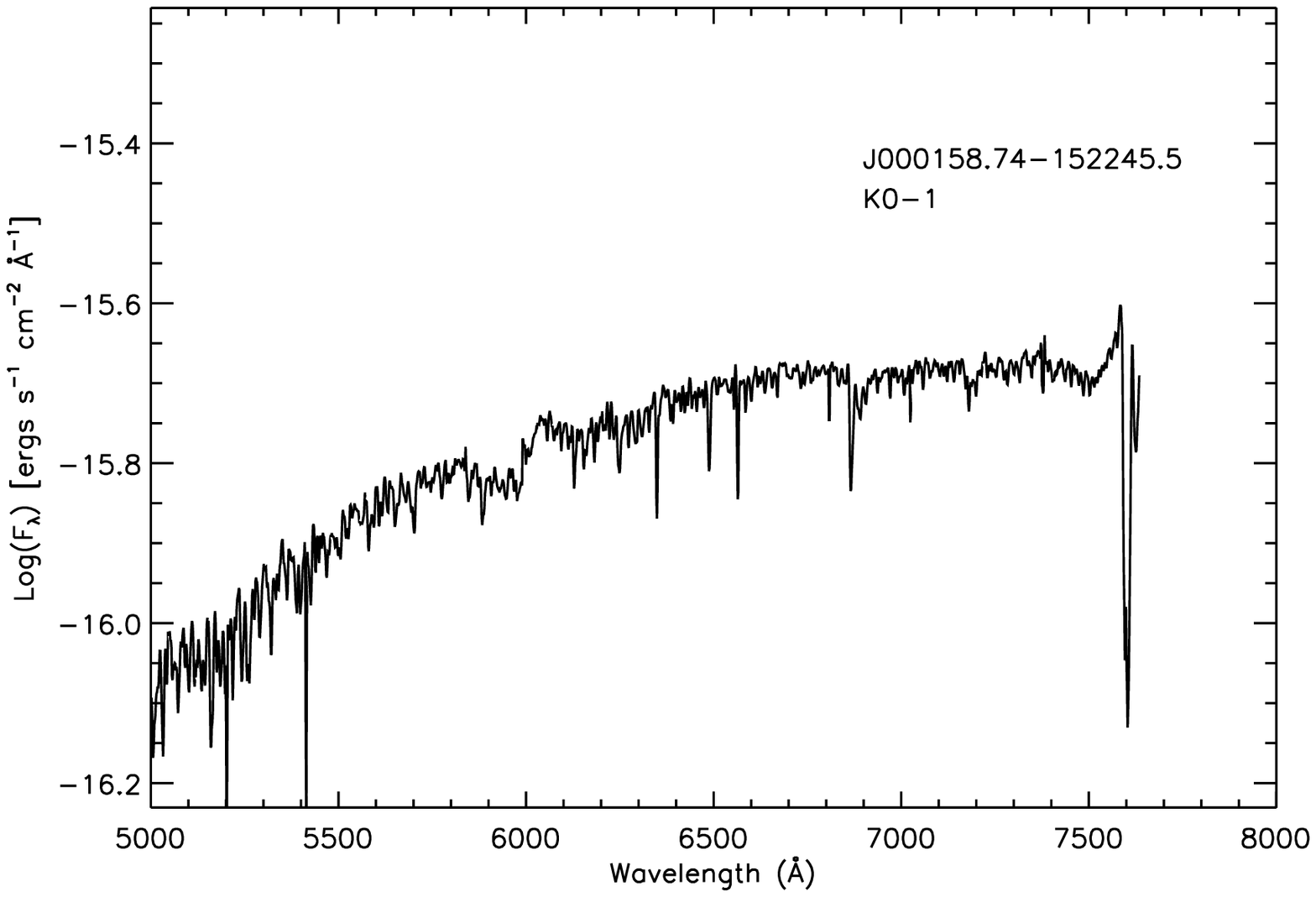}
\plotone{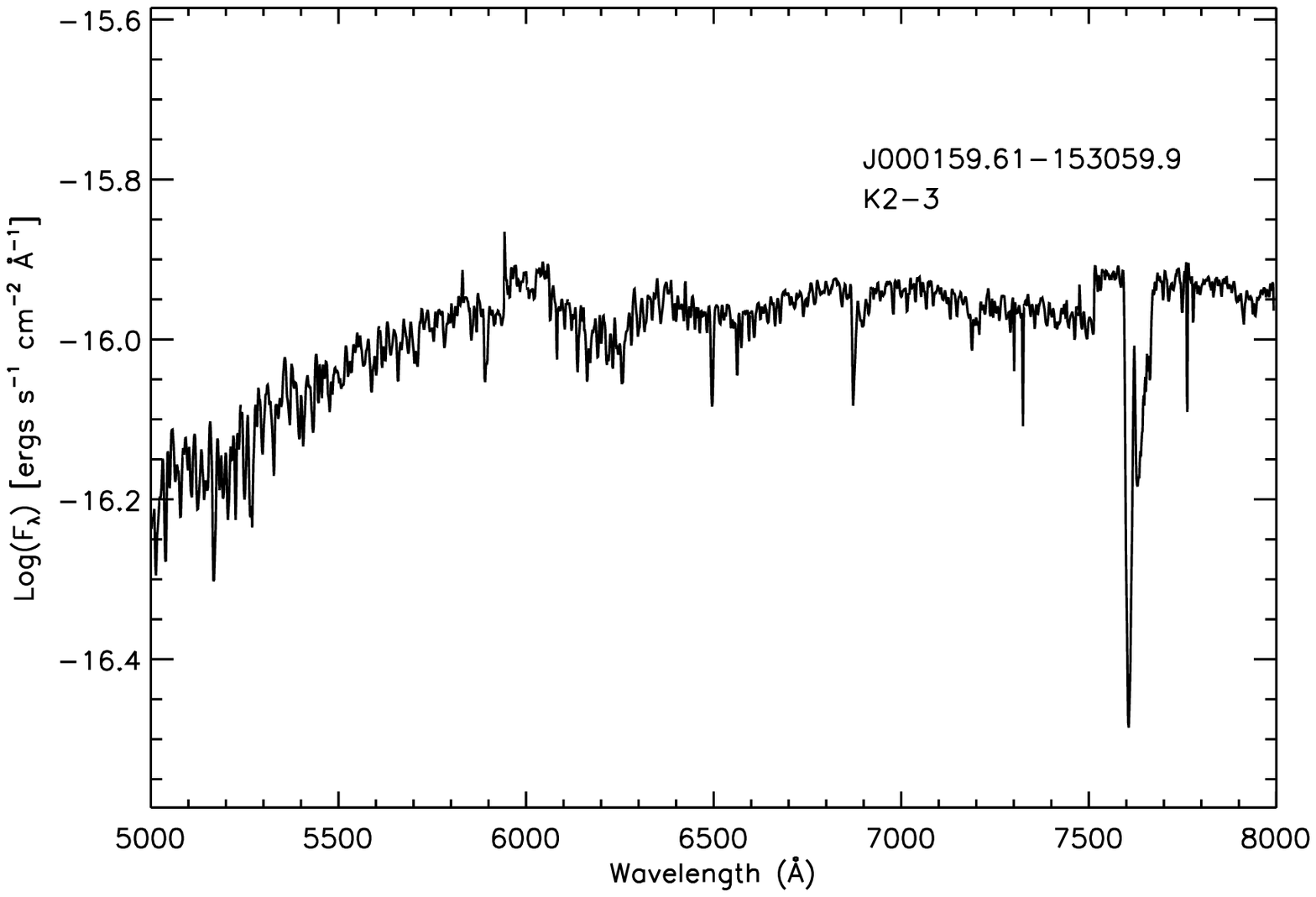}
\plotone{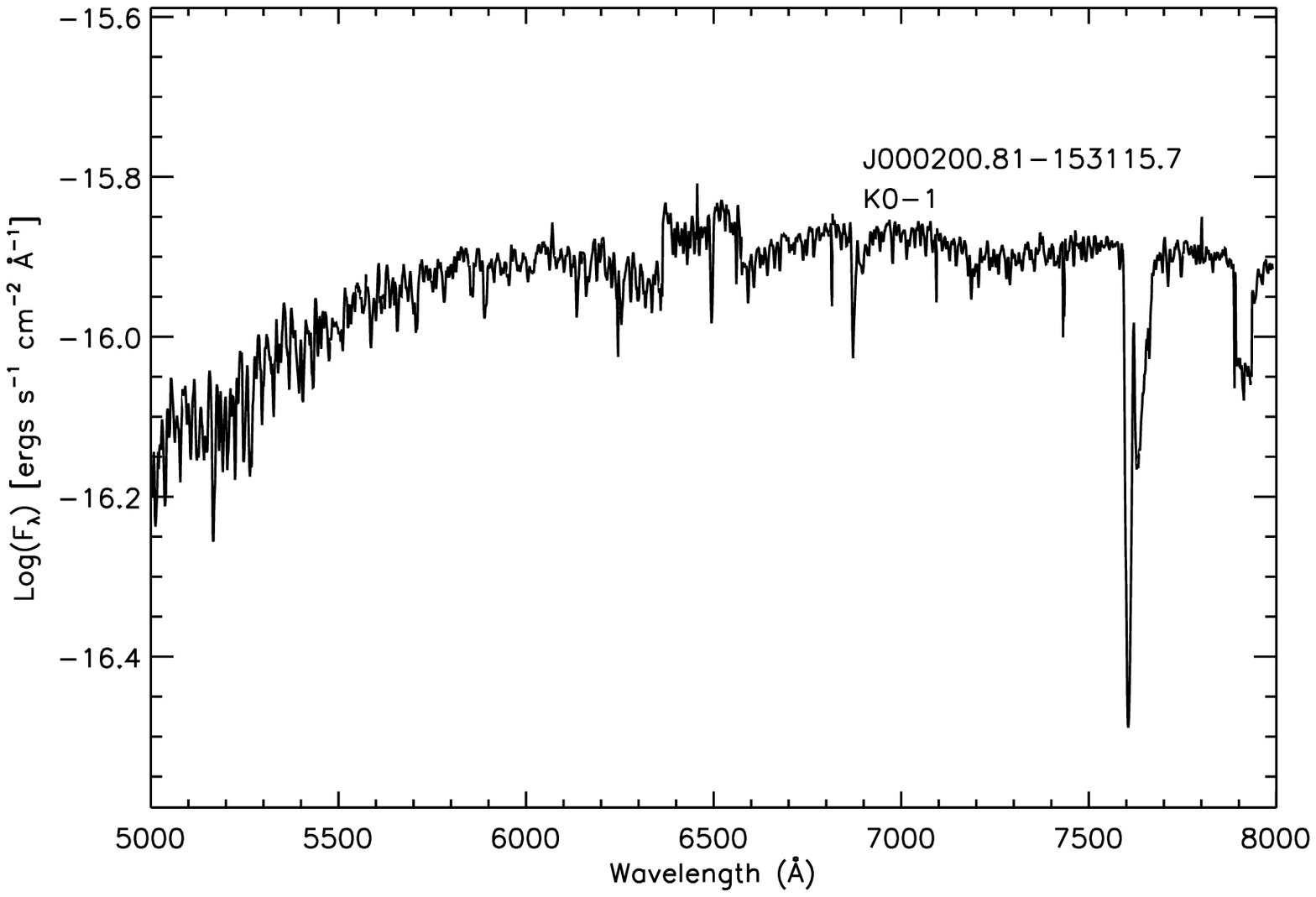}
\plotone{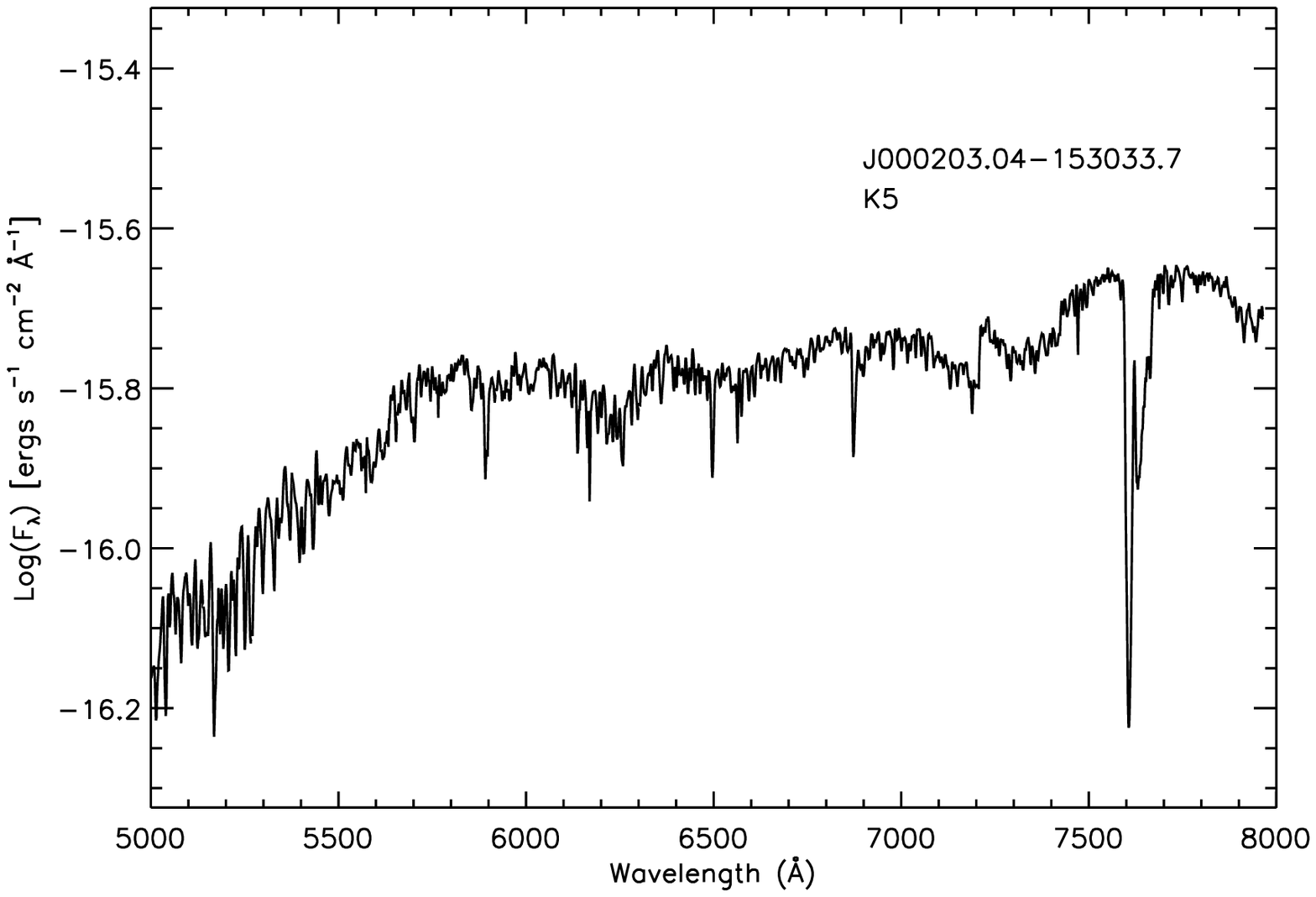}
\caption{Spectrophotometry and spectral types for our 11 WLM RSGs. The strong feature at 7600 \AA\ is the telluric A band.}
\end{figure}

\begin{figure}
\epsscale{0.36}
\plotone{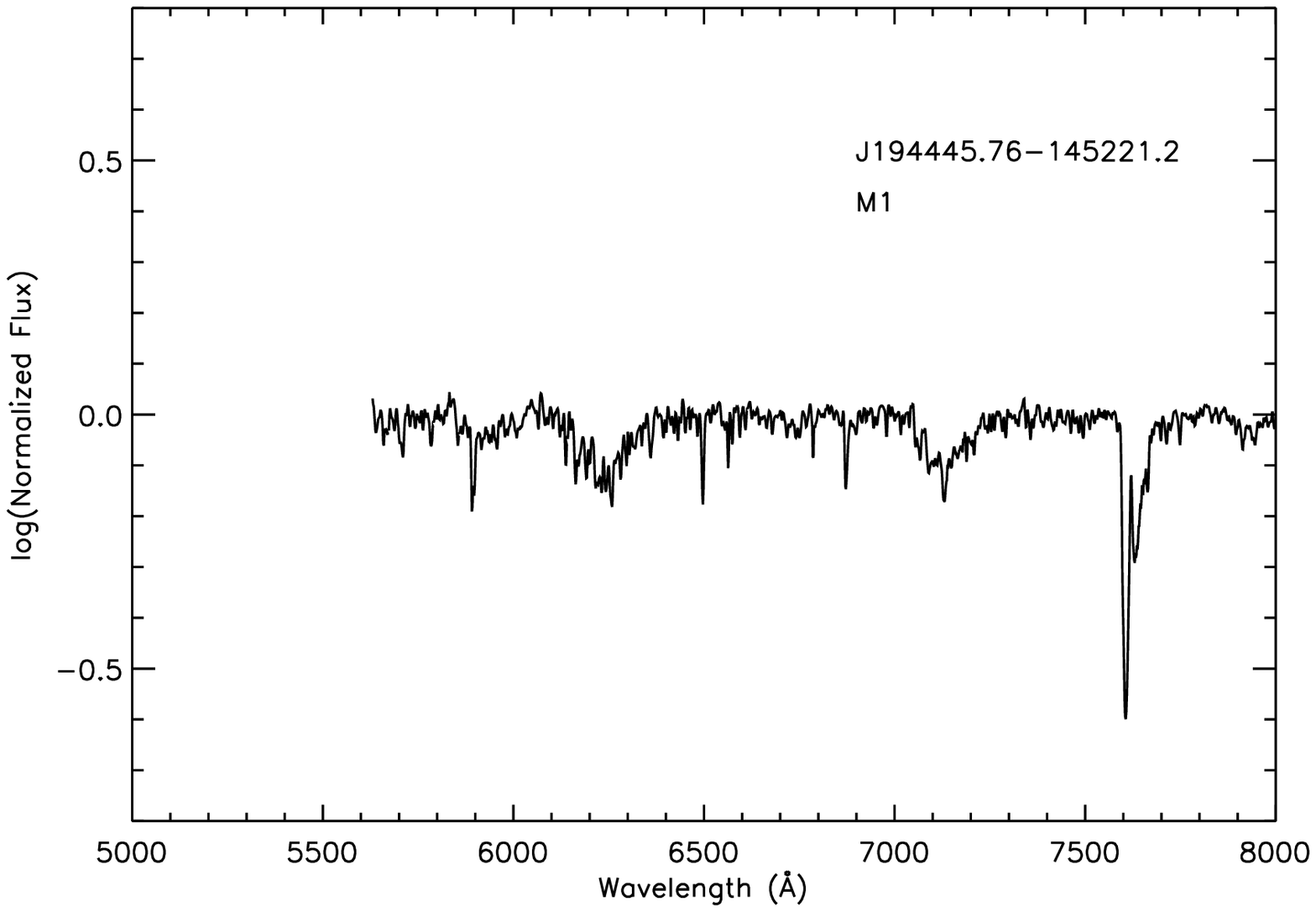}
\plotone{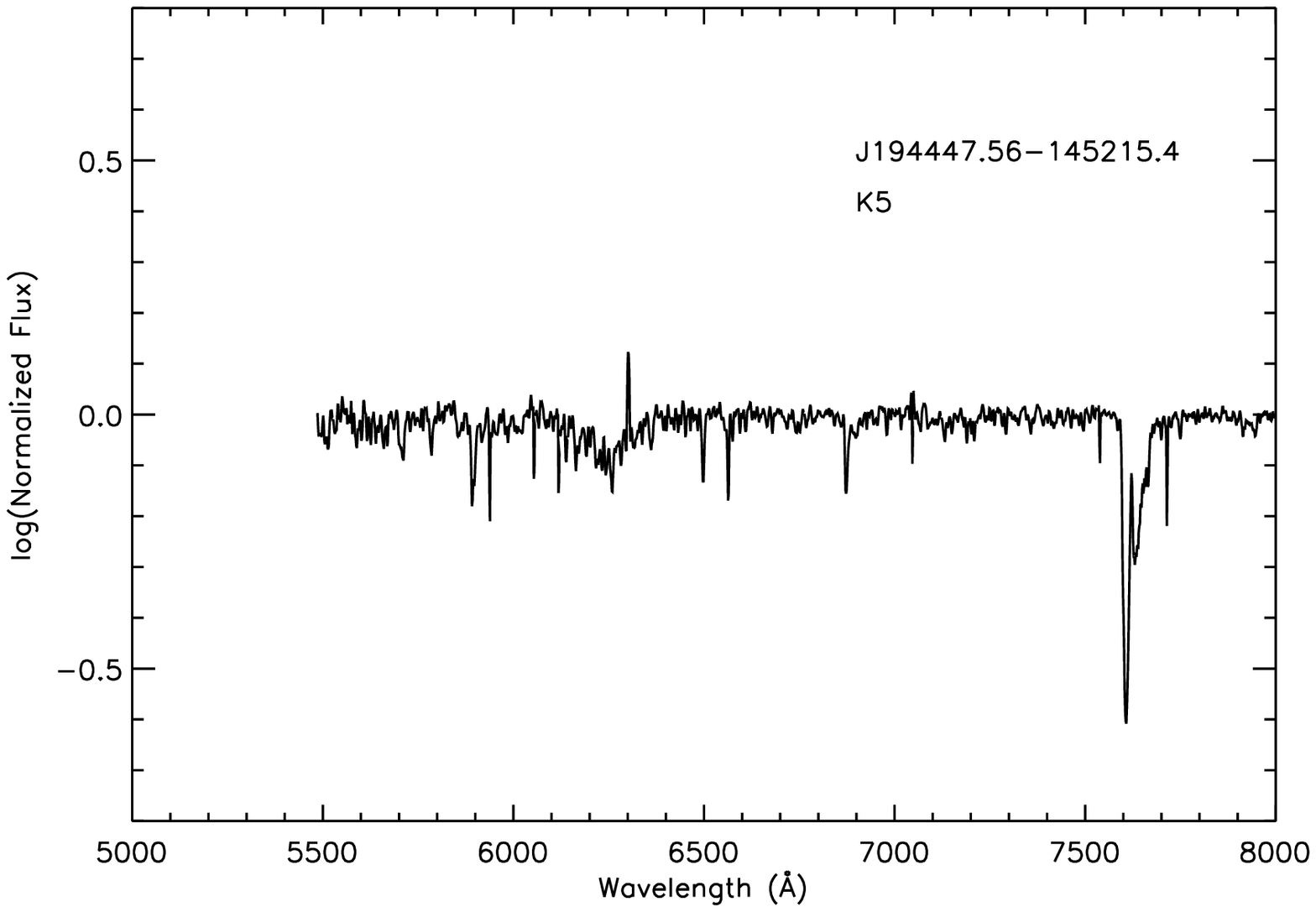}
\plotone{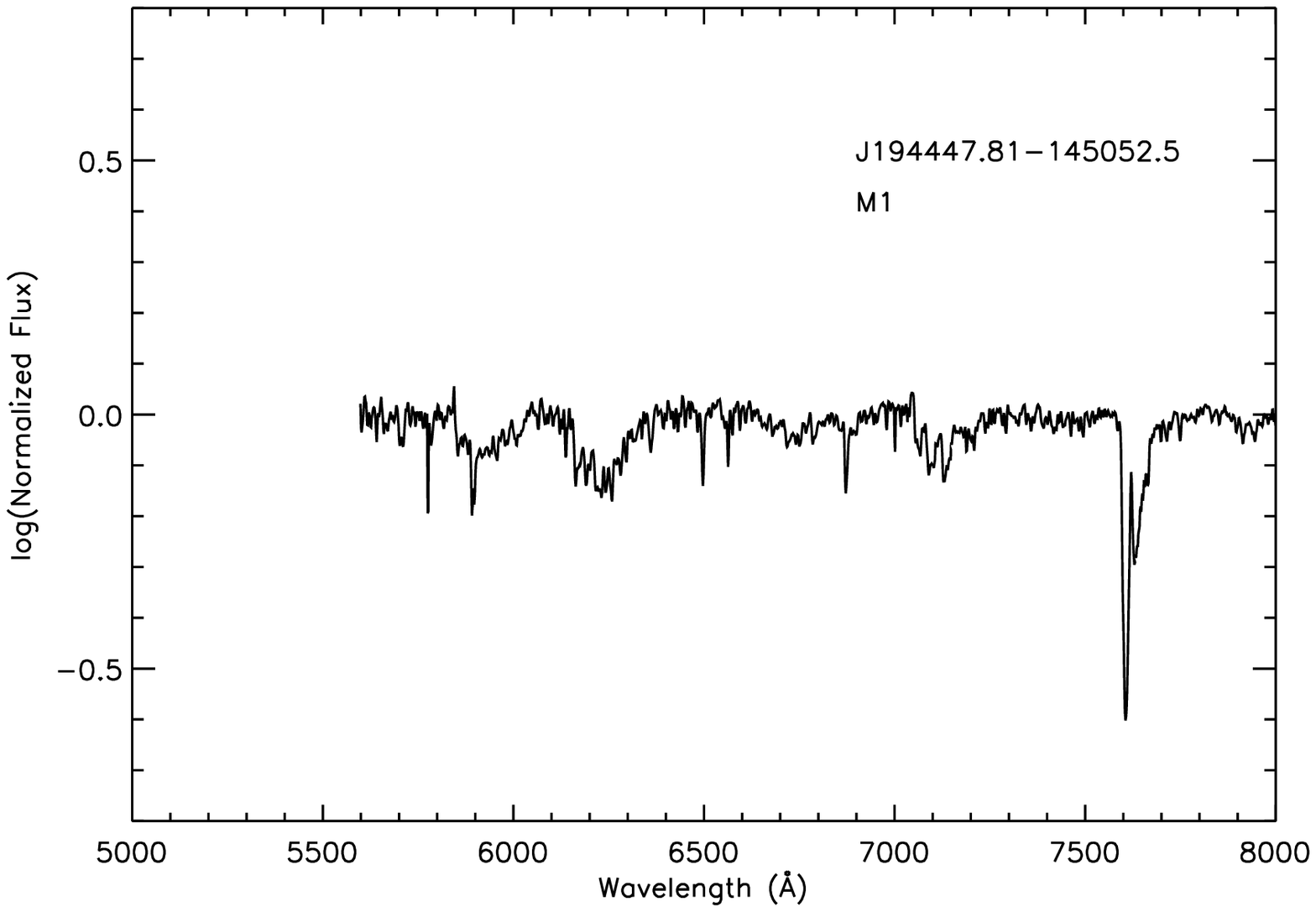}
\plotone{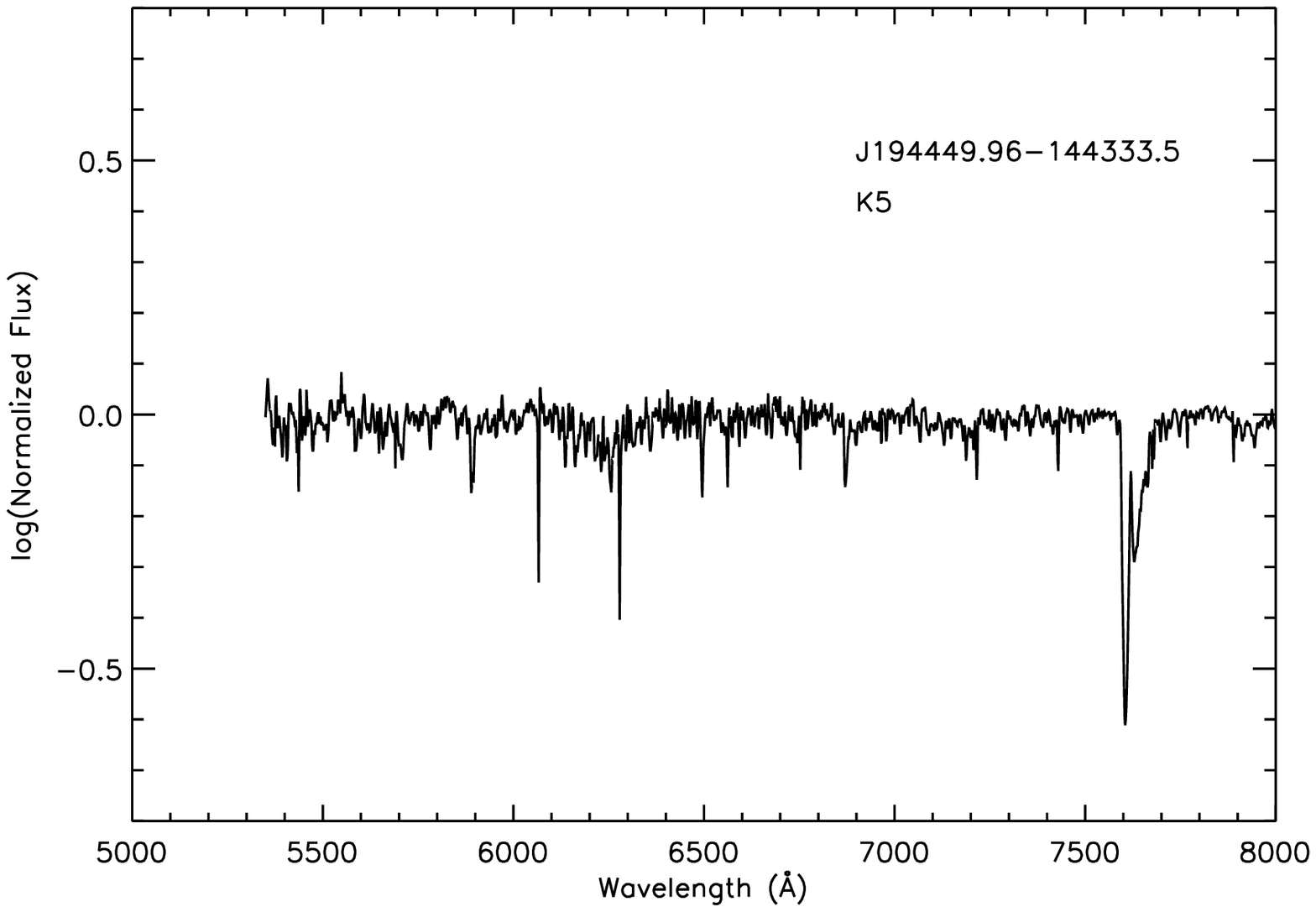}
\plotone{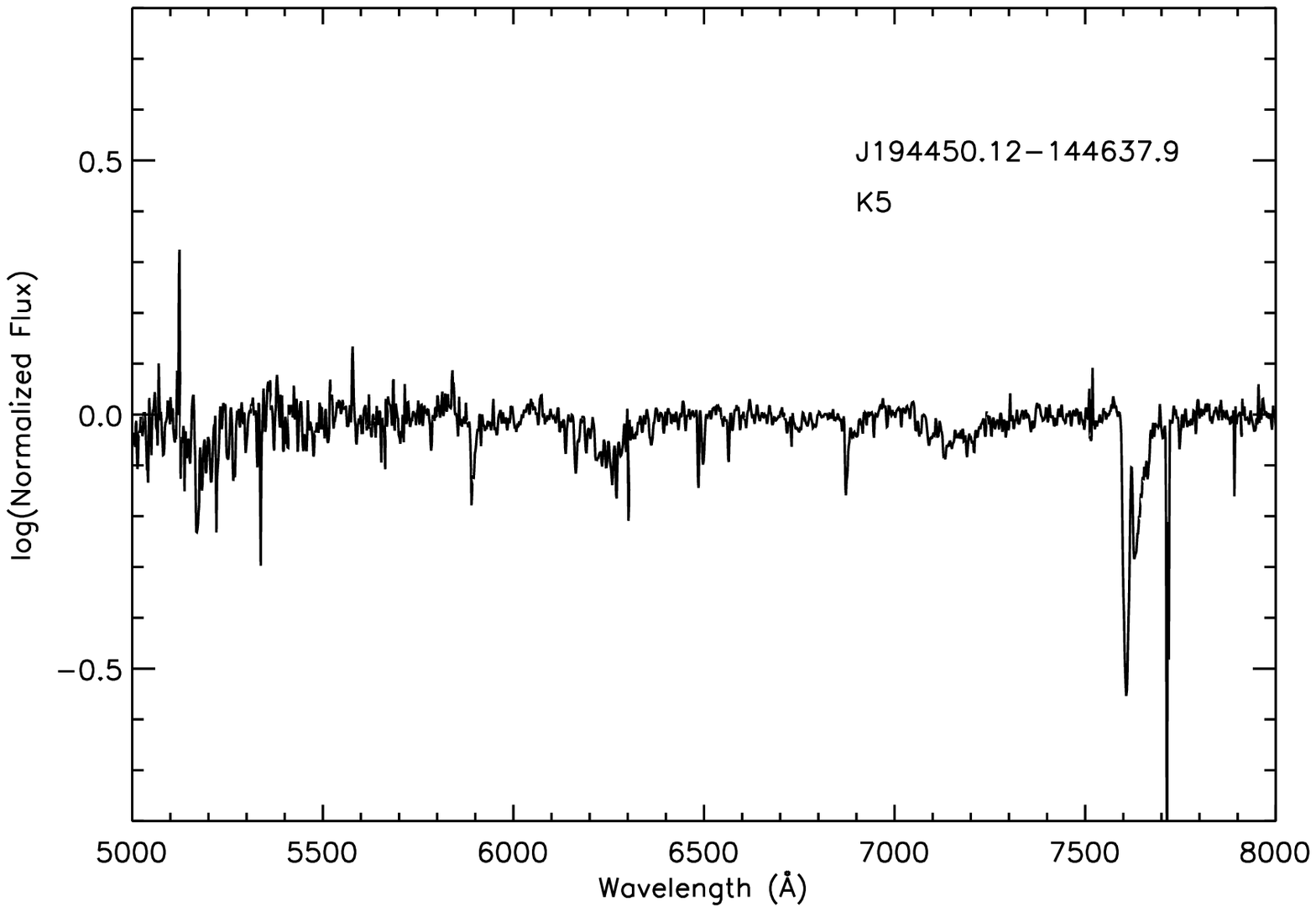}
\plotone{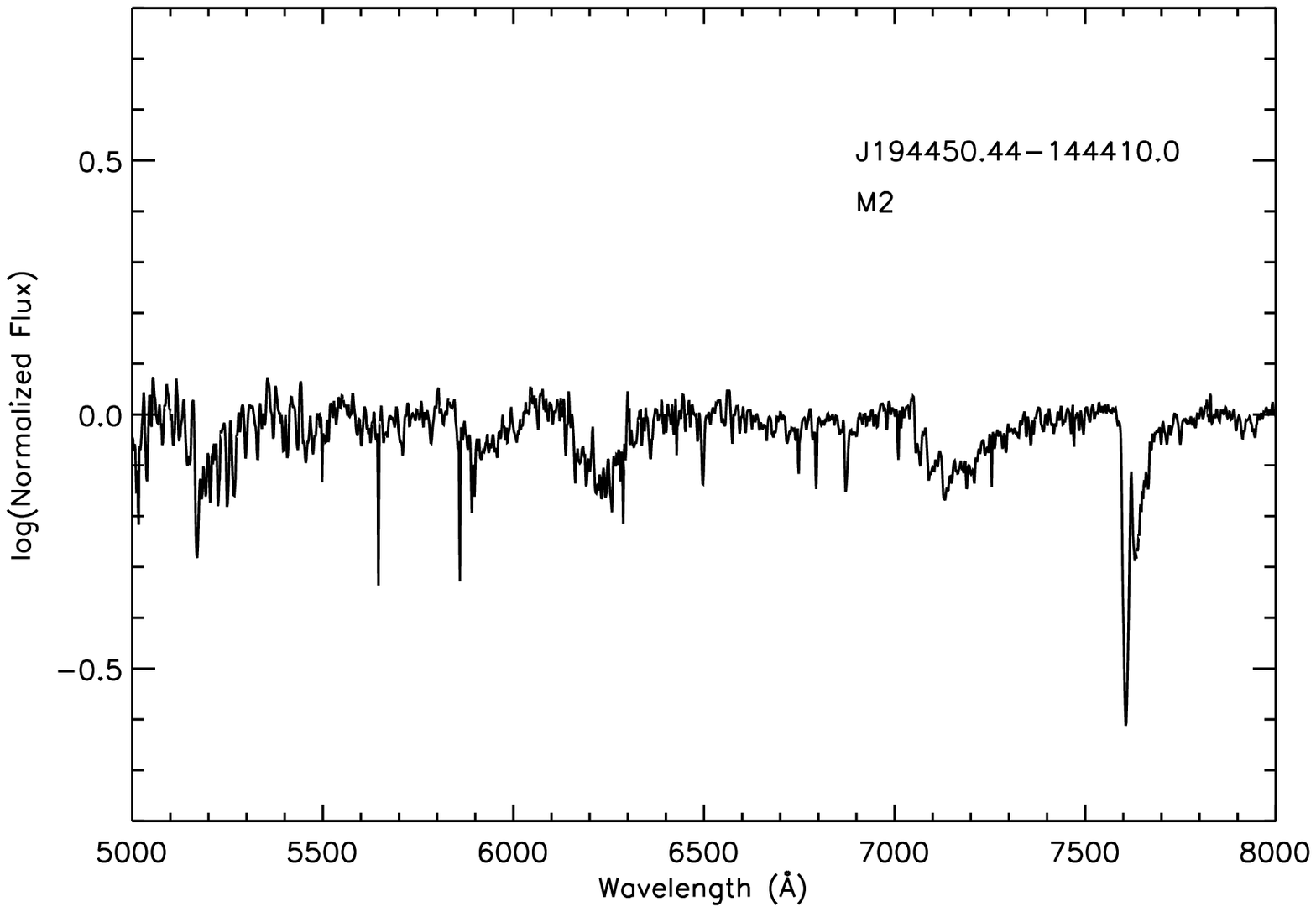}
\plotone{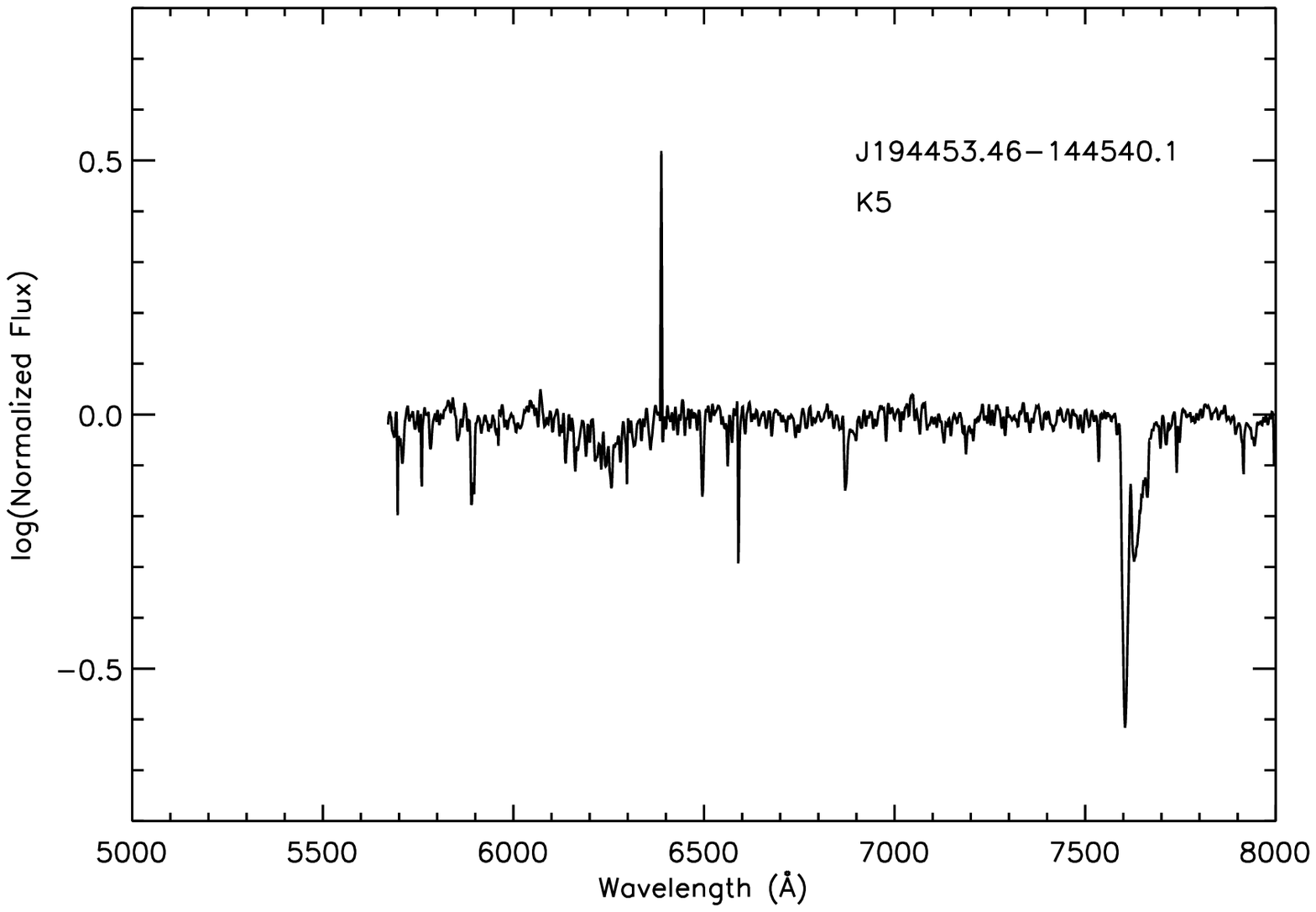}
\plotone{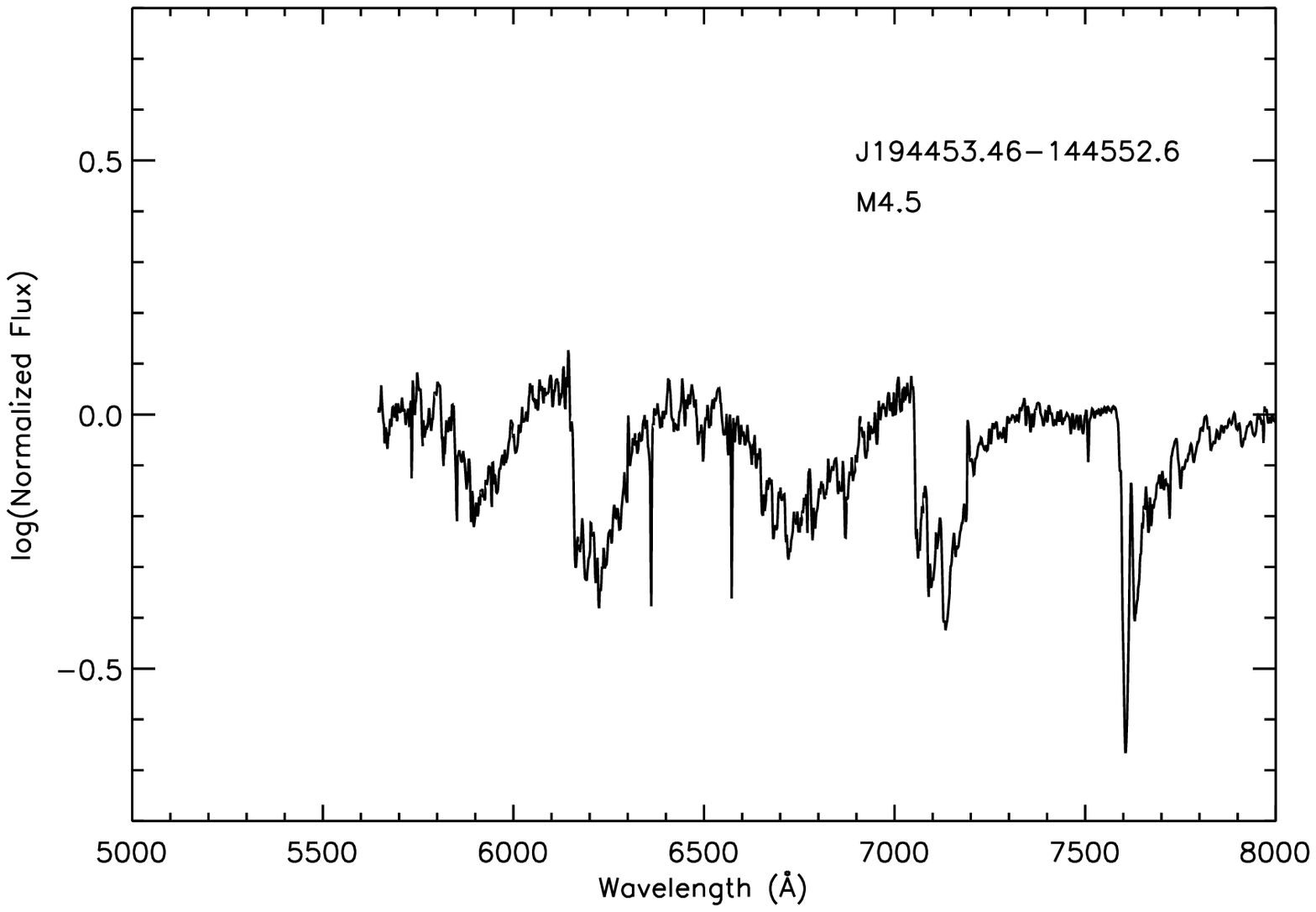}
\plotone{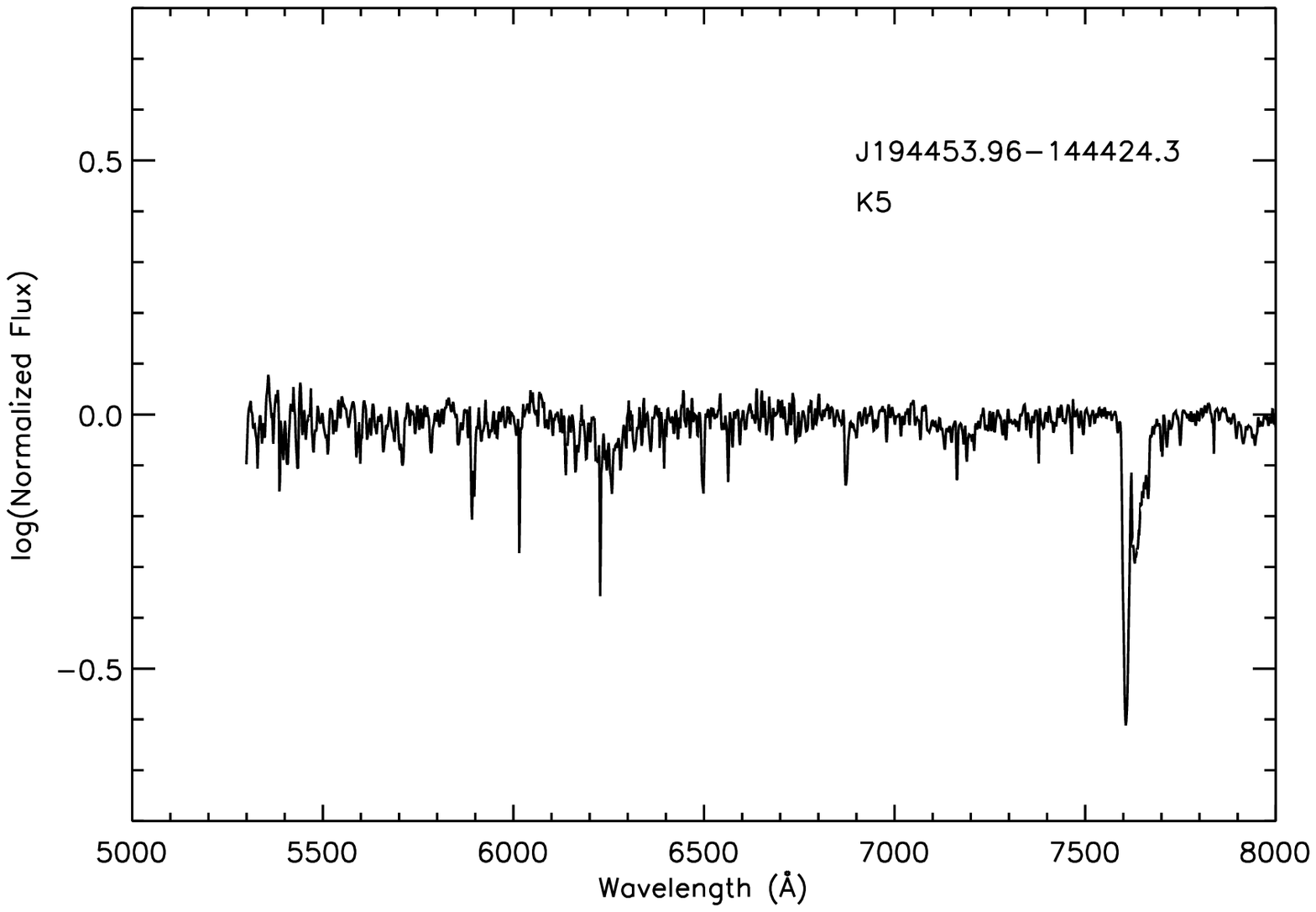}
\plotone{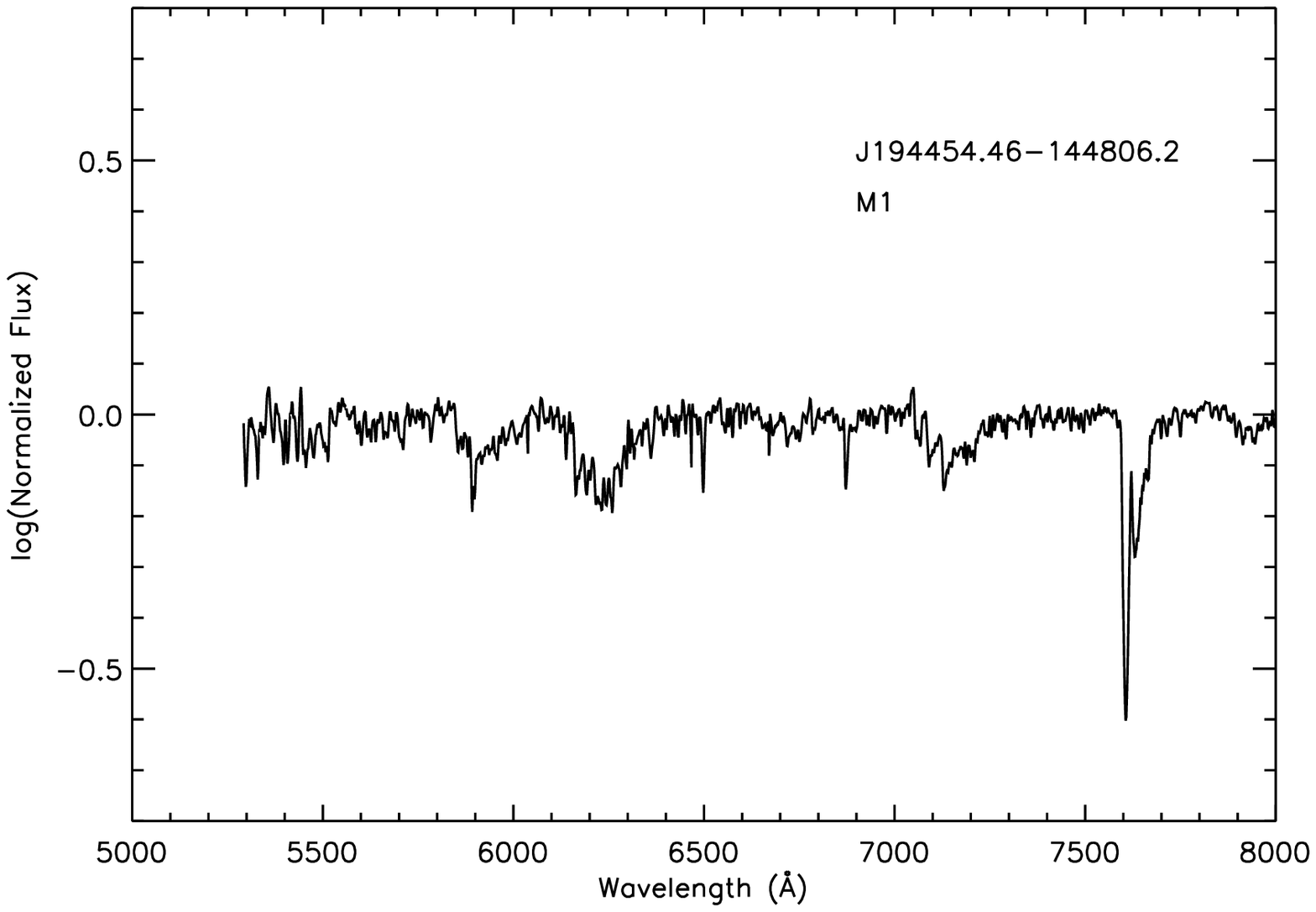}
\plotone{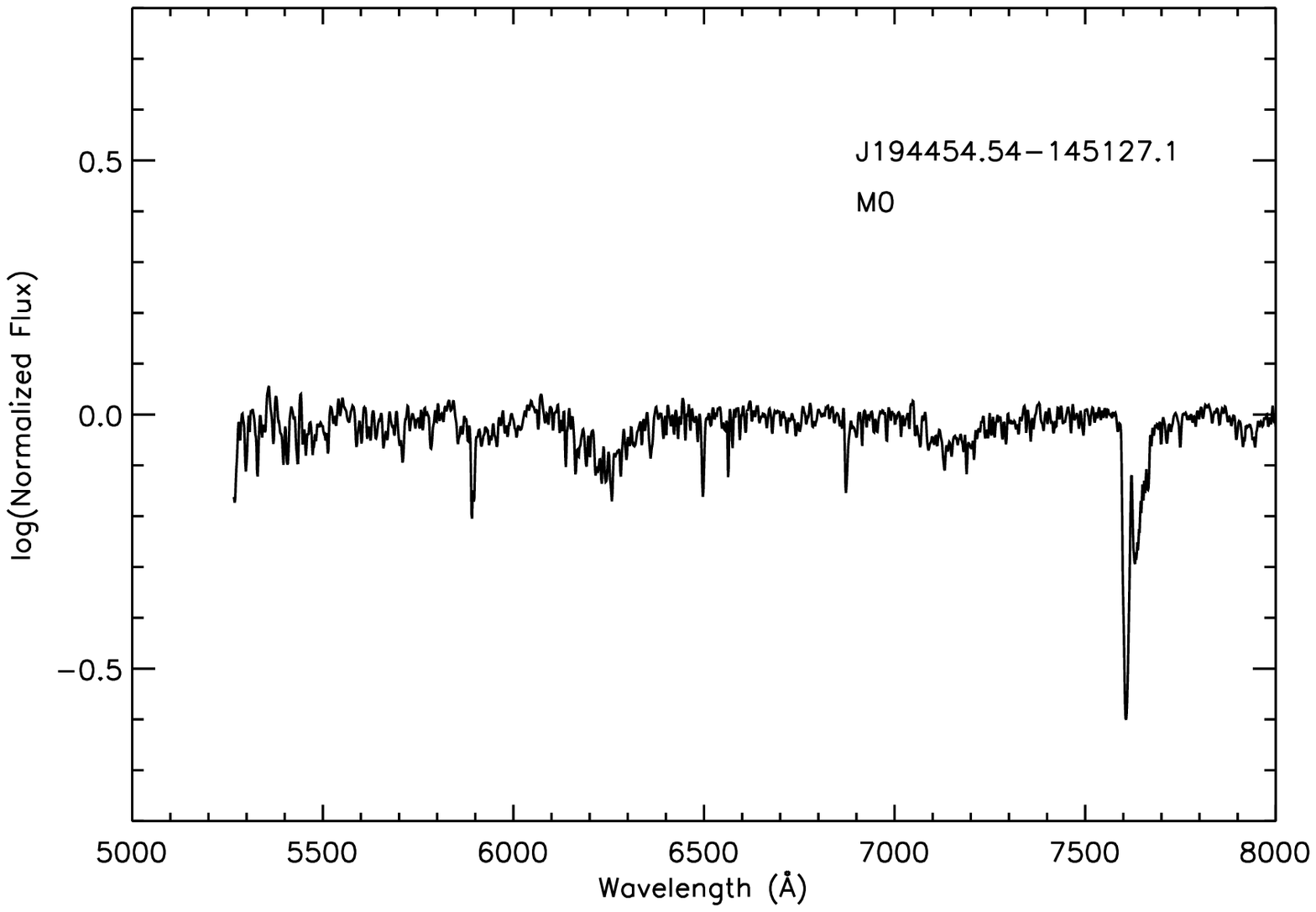}
\plotone{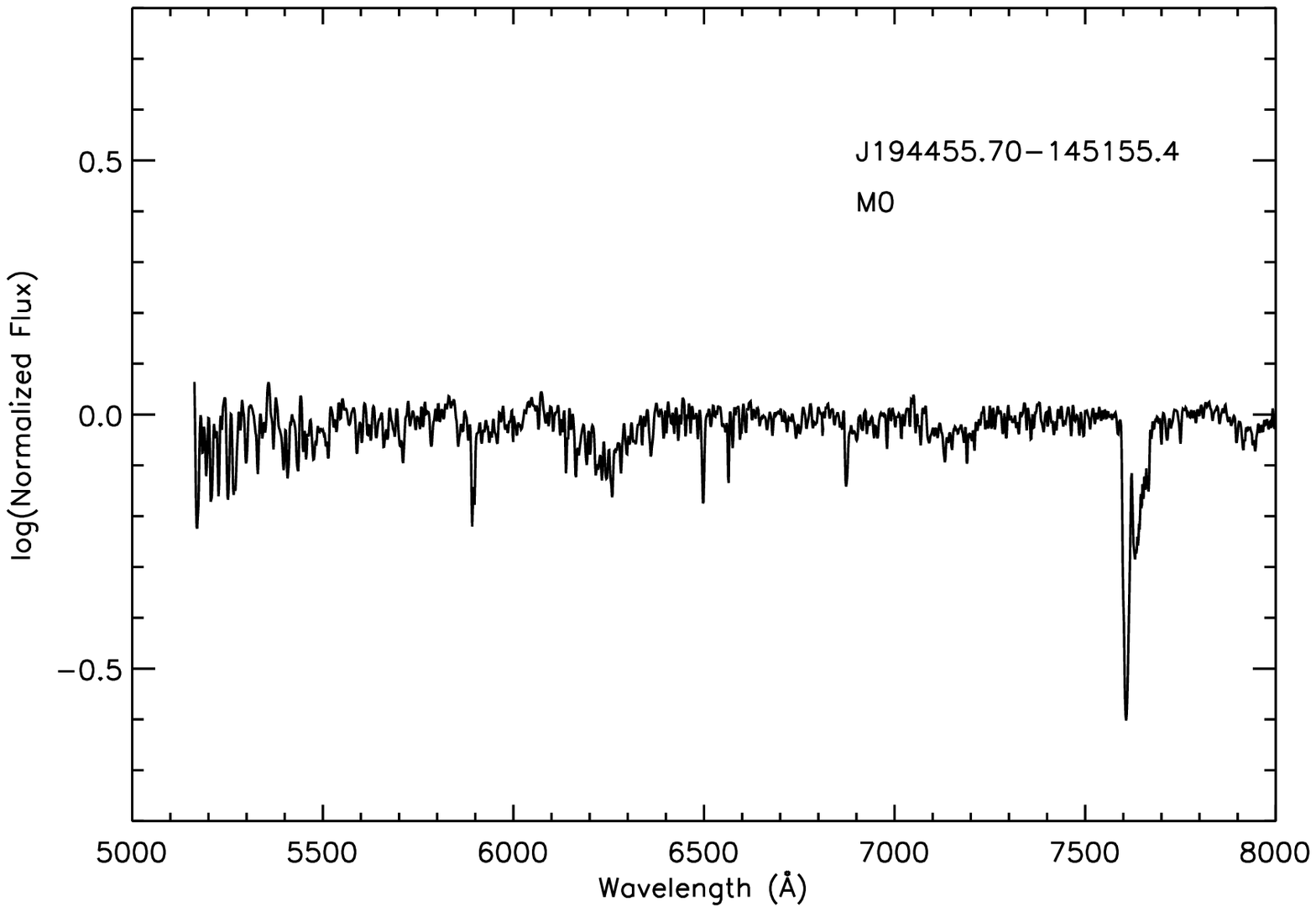}
\plotone{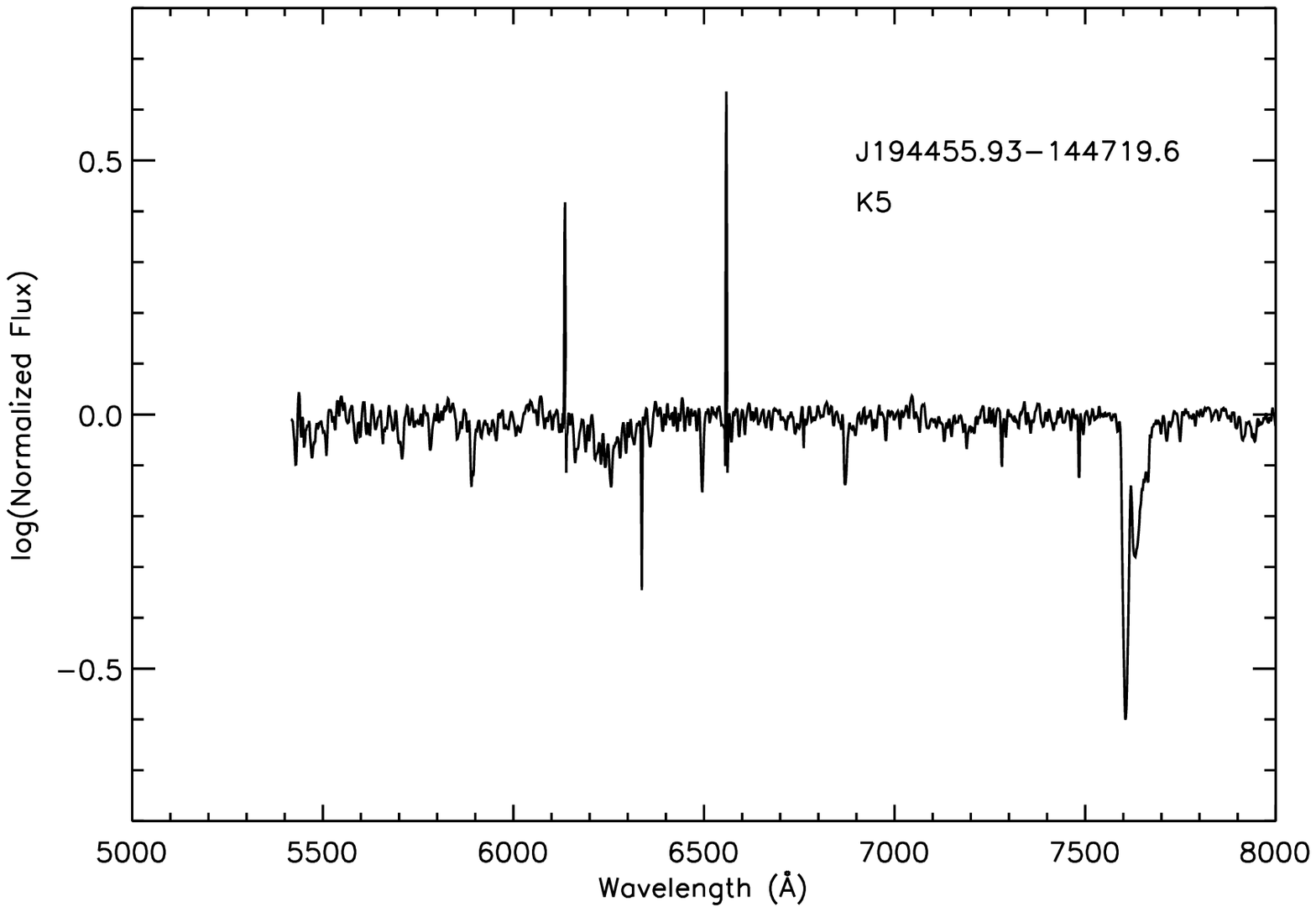}
\plotone{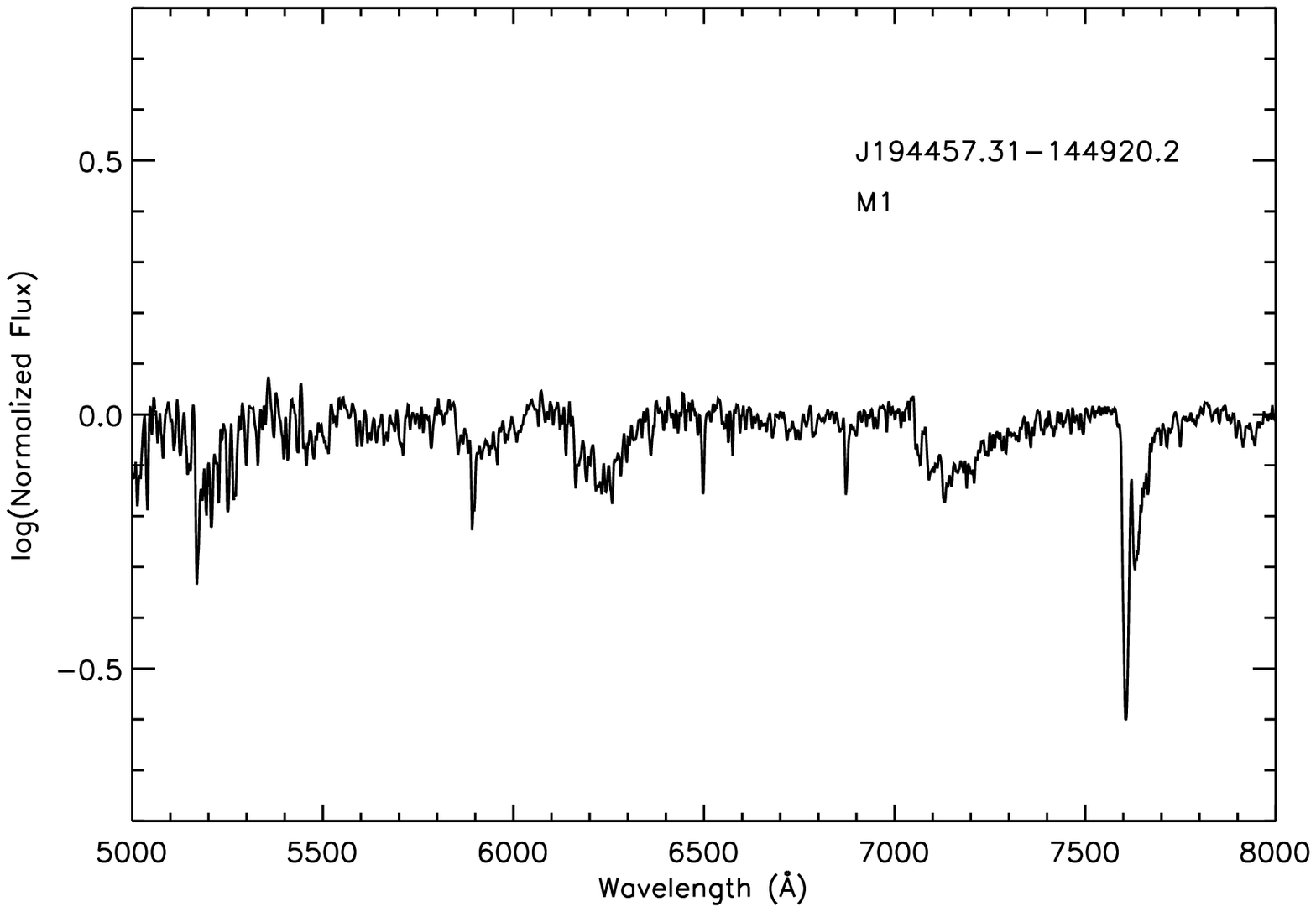}
\plotone{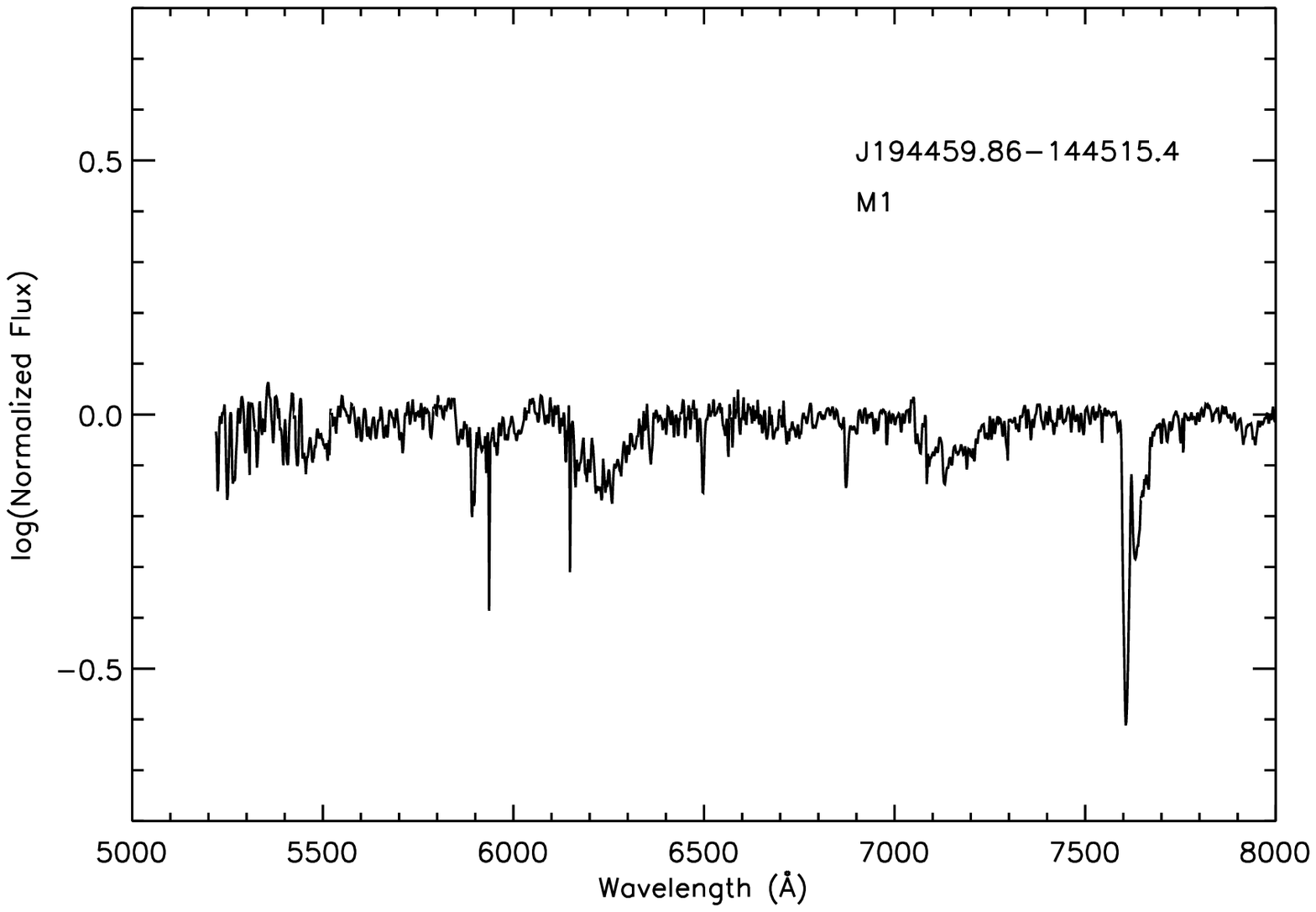}
\plotone{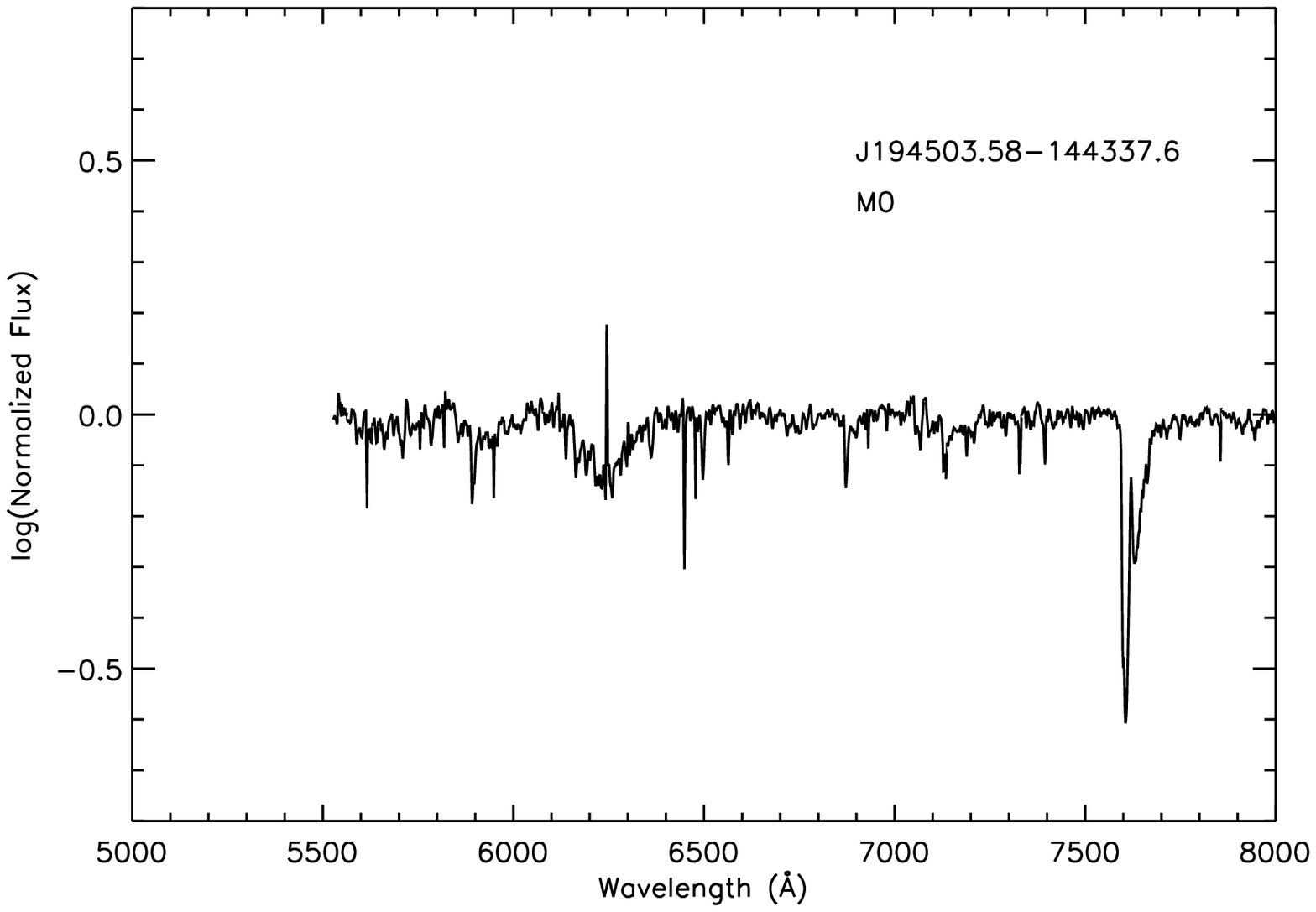}
\caption{Normalized spectroscopy and spectral types for our 16 NGC 6822 RSGs. The strong feature at 7600 \AA\ is the telluric A band.}
\end{figure}

\begin{figure}
\epsscale{1.0}
\plotone{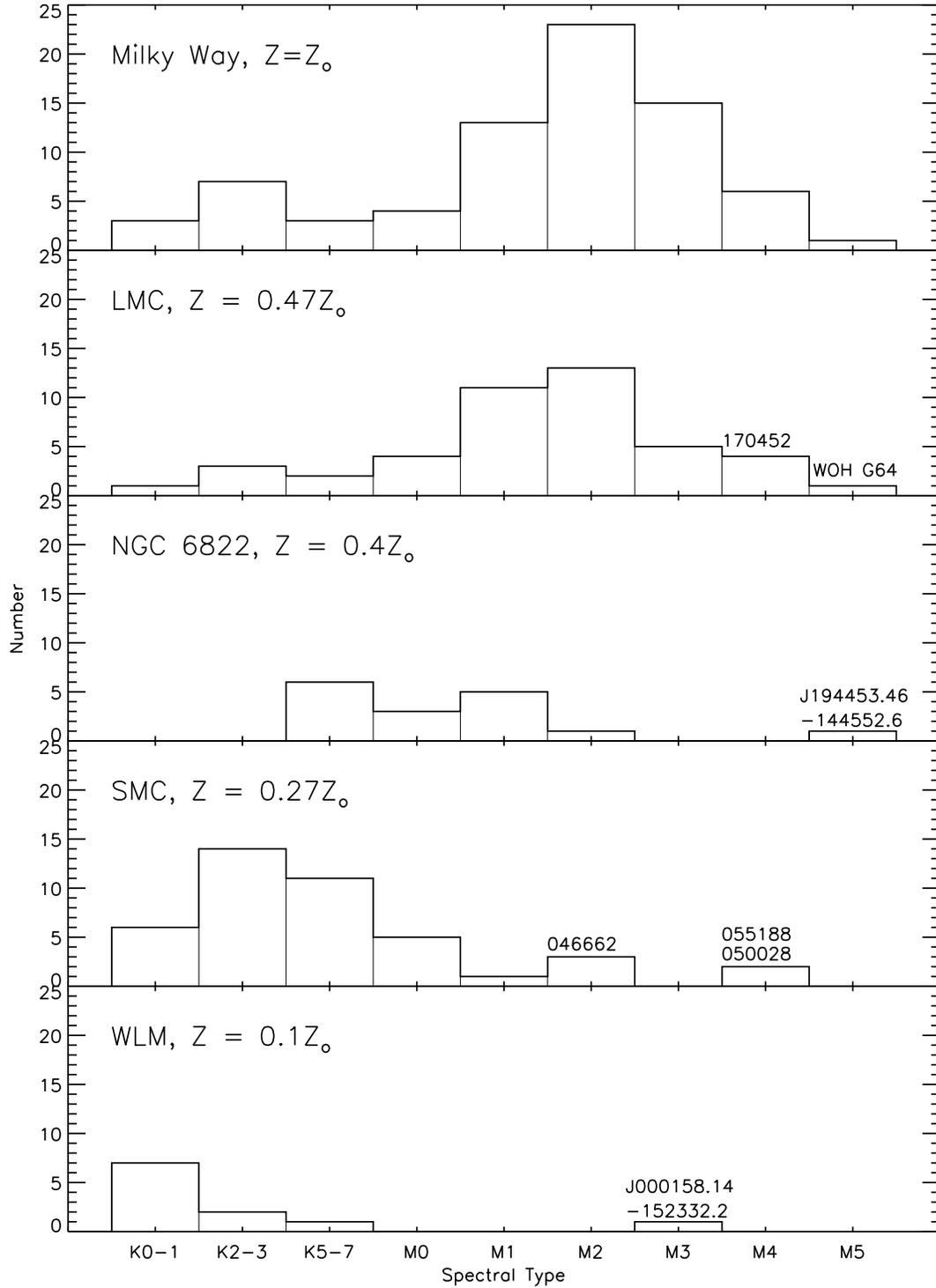}
\vspace{-10pt}
\caption{Histograms of RSG spectral types found in five different Local Group galaxies, plotted from top to bottom in order of decreasing metallicity. Data for the Milky Way RSGs are taken from Levesque et al.\ (2005); for the Magellanic Clouds spectral type data comes from Levesque et al.\ (2006, 2007) and Massey et al.\ (2007b). Spectral types for NGC 6822 and WLM RSGs are from this work. In this comparison we can observe the progression of the dominant spectral type towards earlier types at lower metallicities that was previously discussed by Elias et al.\ (1985) and Massey \& Olsen (2003). Late-type RSGs from Levesque et al.\ (2007) and this work are labeled.}
\end{figure}

\begin{figure}
\epsscale{0.90}
\plotone{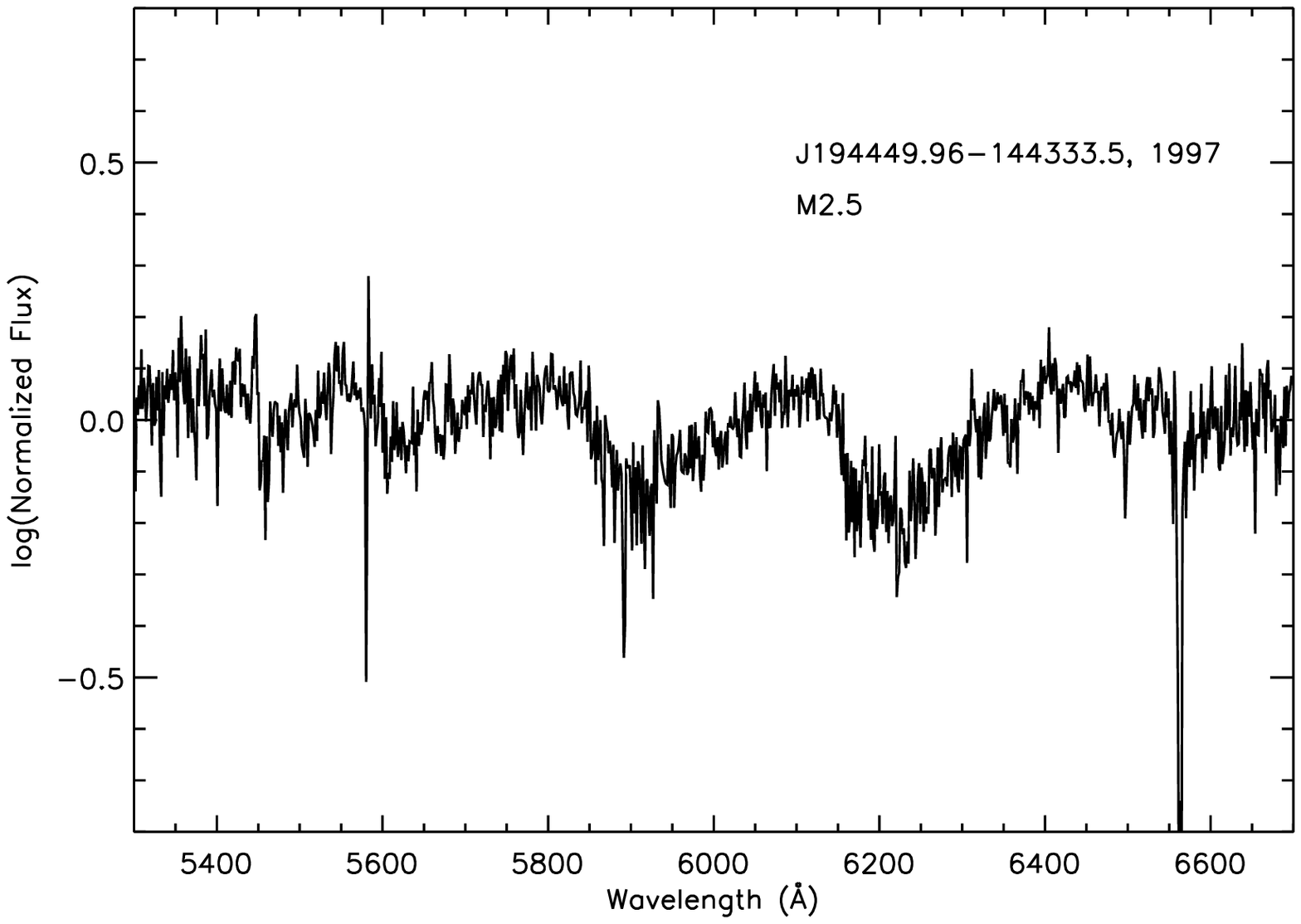}
\plotone{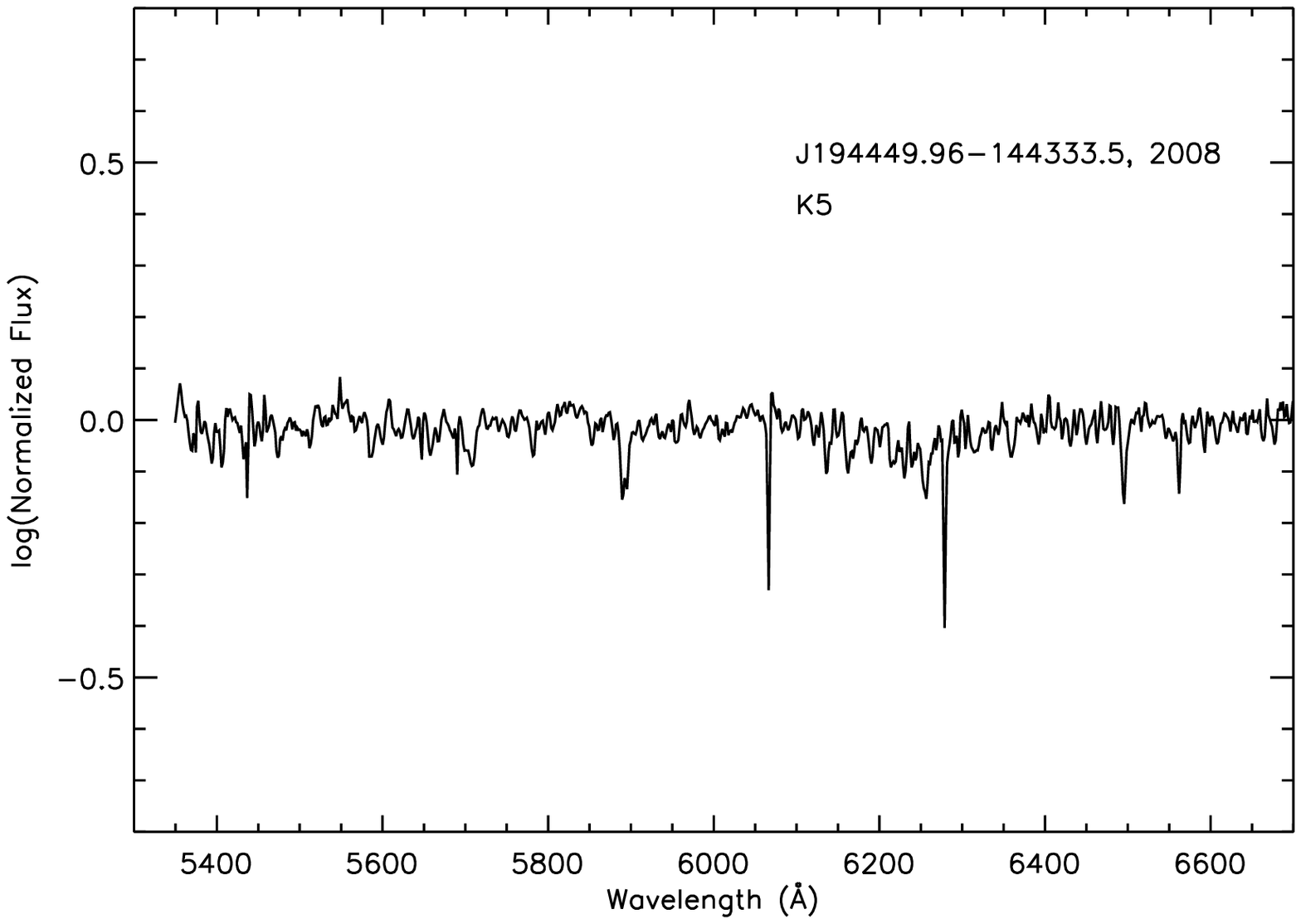}
\caption{Normalized spectra of the NGC 6822 RSG J194449.96-144333.5 from 1997 observations described in Massey (1998; top) and this work (bottom). The 1997 spectrum shows strong 5448\AA, 5847\AA, and 6158\AA\ TiO absorption bands, consistent with a spectral type of M2.5, while the 2008 spectrum shows only the weak 6158\AA\ TiO band, consistent with a K5 spectral type.}
\end{figure}

\begin{figure}
\epsscale{0.65}
\plotone{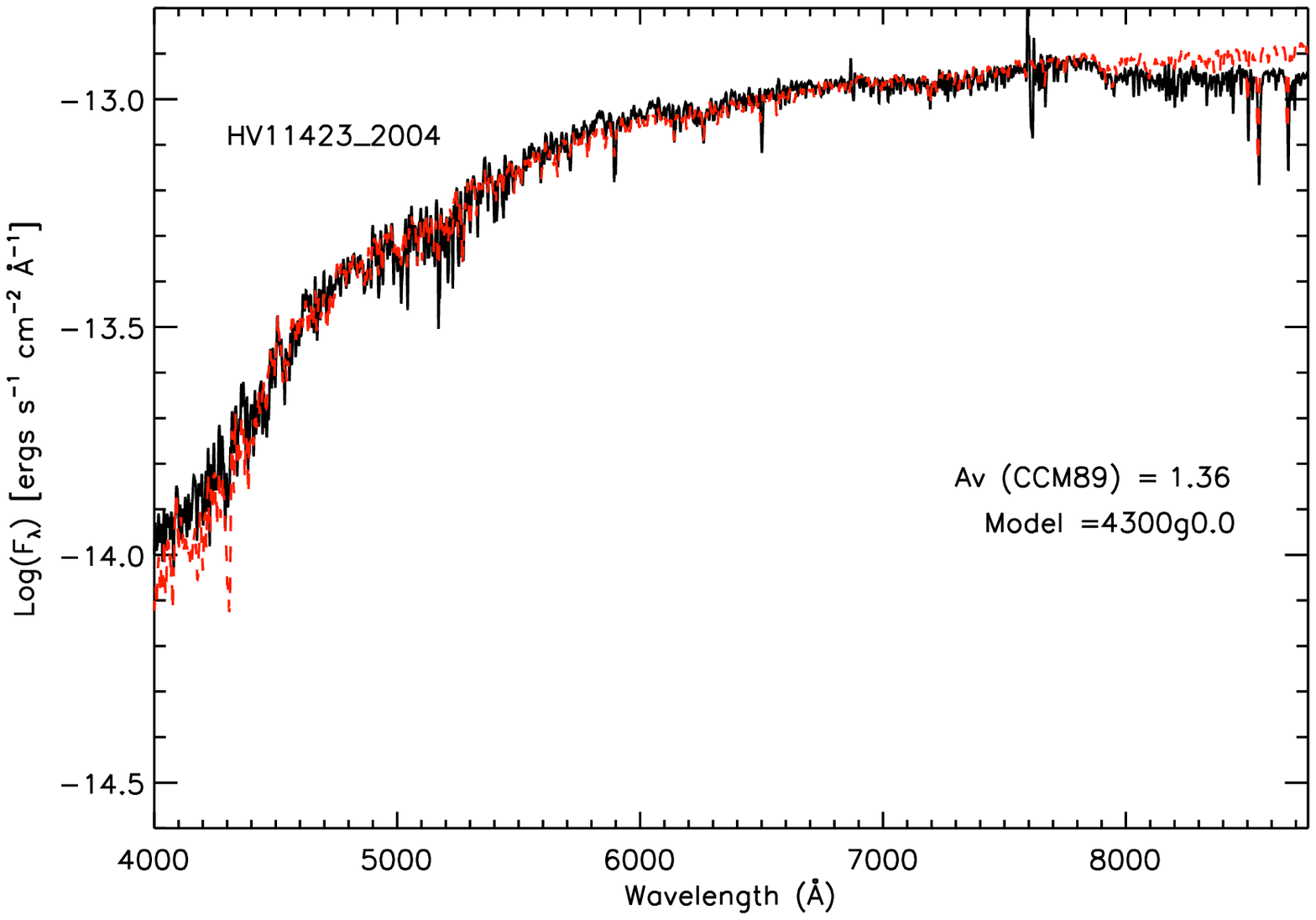}
\plotone{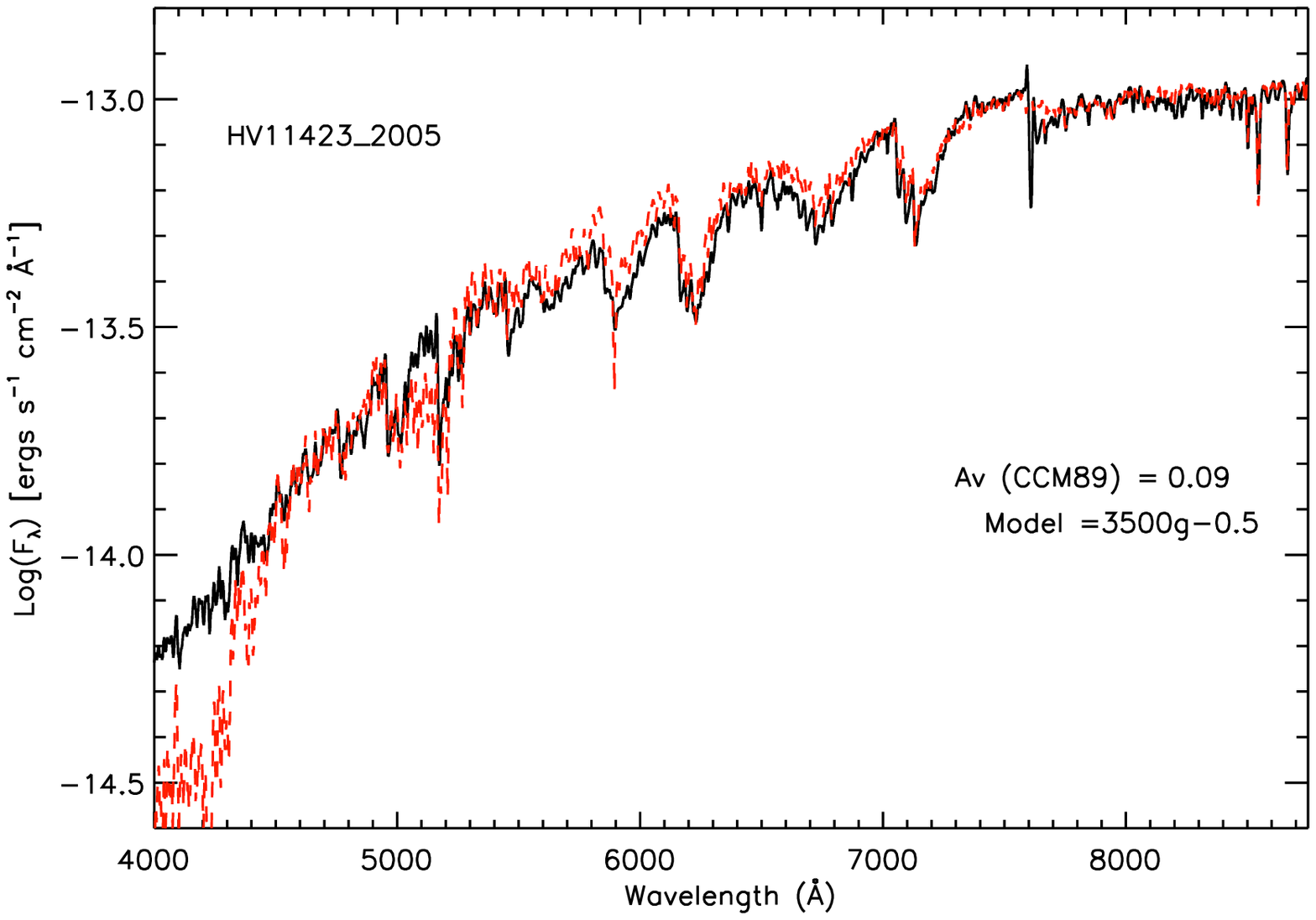}
\plotone{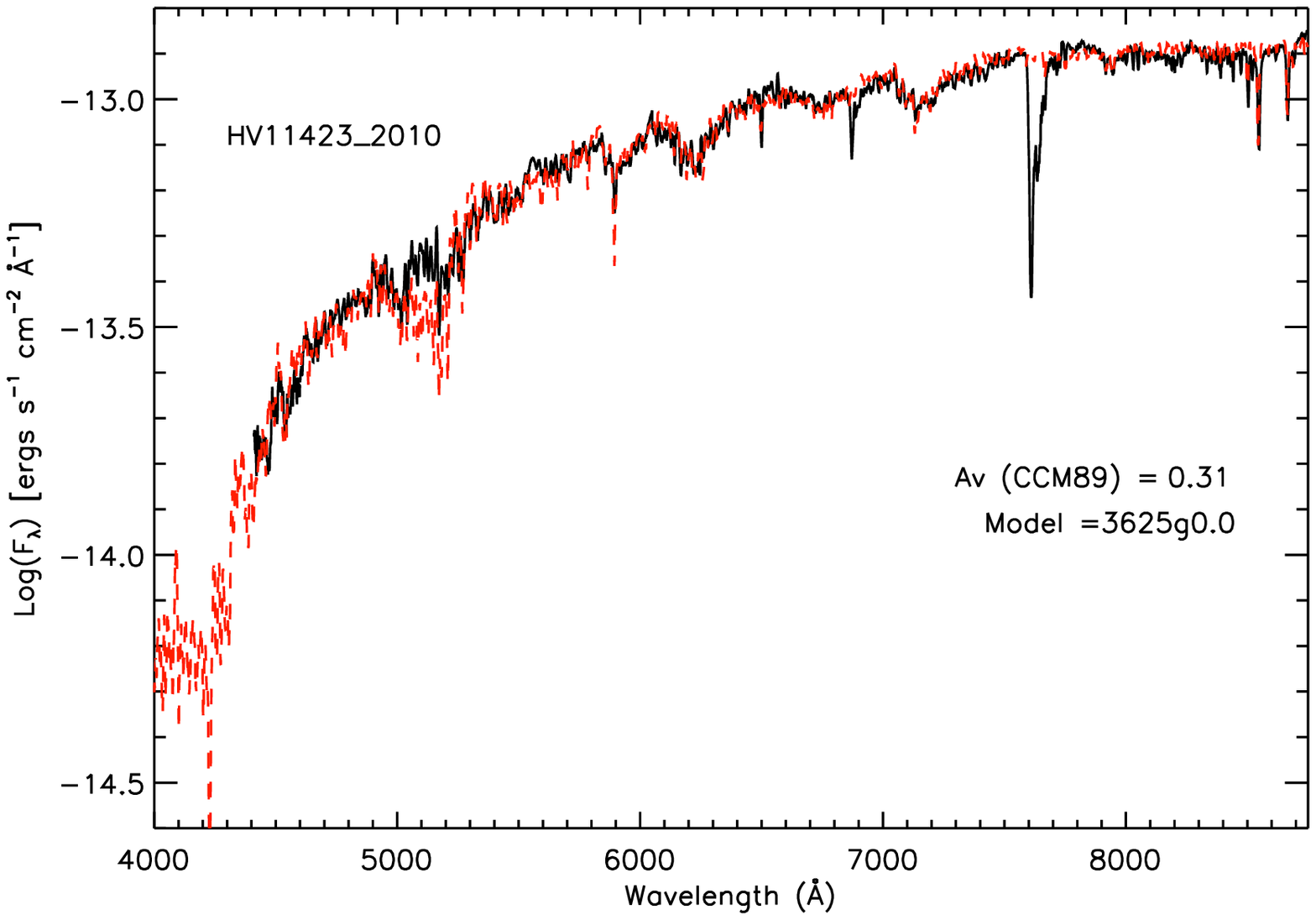}
\vspace{-10pt}
\caption{Moderate-resolution spectra of the unusual variable SMC 050028 in 2004 (top), 2005 (center), and 2010 (bottom). Observed spectra are shown in black; MARCS stellar atmosphere models (Plez 2003; Gustafsson et al.\ 2003, 2008) at SMC metallicity are shown in red. The variation in the star's spectral type and $T_{\rm eff}$ is apparent from the changing strengths of the TiO bands; using the MARCS models we determine a $T_{\rm eff}$ of 4300 K from the 2004 data, 3500 K in 2005, and 3625 K in 2010.}
\end{figure}

\end{document}